\documentclass[12pt,english,floatfix,nofootinbib,superscriptaddress,aps,prd,preprint]{revtex4-2}
\usepackage[utf8]{inputenc}       
\usepackage[english]{babel}       
\usepackage[utf8]{inputenc}
\usepackage{textcomp}
\usepackage{amsmath,amsthm,amssymb,amsfonts,mathrsfs,amsbsy} 
\usepackage{tensor}               
\usepackage{slashed}  
\usepackage{bbm}
\usepackage{esint}    
\usepackage[a4paper, margin=1.6cm]{geometry}
\usepackage{cancel}    
\usepackage{caption}
\usepackage{graphicx}             
\usepackage{natbib}
\usepackage{float}                
\usepackage{subfig}               
\usepackage[font=small,labelfont=bf]{caption} 
\usepackage{multirow}             
\usepackage{array}                
\usepackage{tikz}                 
\usetikzlibrary{quotes,angles,arrows,decorations.markings} 
\usepackage{makecell} 

\usepackage{lipsum}

\usepackage{xcolor}               
\usepackage{color}                
\usepackage{textcomp}             
\usepackage{units}                

\newcommand{\be}{\begin{equation}}
\newcommand{\ee}{\end{equation}}
\newcommand{\Be}{\begin{eqnarray}}
\newcommand{\Ee}{\end{eqnarray}}

\newcommand{\mincir}{\raise
-3.truept\hbox{\rlap{\hbox{$\sim$}}\raise4.truept\hbox{$<$}\ }}
\newcommand{\magcir}{\raise
-3.truept\hbox{\rlap{\hbox{$\sim$}}\raise4.truept\hbox{$>$}\ }}

\newcolumntype{Y}{>{\centering\arraybackslash}X}
\providecommand{\U}[1]

\usepackage[dvips]{epsfig}        
\usepackage[dvips]{graphicx}      
\usepackage{hyperref}             
\hypersetup{
    colorlinks=true,              
    breaklinks=true,              
    citecolor=blue,               
    linkcolor=[rgb]{0,0.5,0.9},   
    urlcolor=red,                 
    filecolor=green               
}

\newcommand{\ie}{\begin{equation}}
\newcommand{\fe}{\end{equation}}
\newcommand{\se}{\begin{eqnarray}}
\newcommand{\ff}{\end{eqnarray}}

\begin{document}

\title{Shadows and lensing signatures of a rotating black hole in a Hernquist dark matter halo}


\author{A. A. Ara\'{u}jo Filho}
\email{dilto@fisica.ufc.br}
\affiliation{Departamento de Física, Universidade Federal da Paraíba, Caixa Postal 5008, 58051--970, João Pessoa, Paraíba,  Brazil.}
\affiliation{Departamento de Física, Universidade Federal de Campina Grande Caixa Postal 10071, 58429-900 Campina Grande, Paraíba, Brazil.}
\affiliation{Center for Theoretical Physics, Khazar University, 41 Mehseti Street, Baku, AZ-1096, Azerbaijan.}

\author{Arun Kumar}
\email{arunbidhan@gmail.com}
\affiliation{Centre for Theoretical Physics, Jamia Millia Islamia, New Delhi 110025, India.}

\author{N. Heidari}
\email{heidari.n@gmail.com}
\affiliation{Center for Theoretical Physics, Khazar University, 41 Mehseti Street, Baku, AZ-1096, Azerbaijan.}
\affiliation{School of Physics, Damghan University, Damghan, 3671641167, Iran.}
\affiliation{Departamento de Física, Universidade Federal de Campina Grande Caixa Postal 10071, 58429-900 Campina Grande, Paraíba, Brazil.}


\author{C. F. S. Pereira}
	\email{carlosfisica32@gmail.com}
	\affiliation{Departamento de F\'isica e Qu\'imica, Universidade Federal do Esp\'irito Santo, Av.Fernando Ferrari, 514, Goiabeiras, Vit\'oria, ES 29060-900, Brazil.}

\author{Amilcar R. Queiroz}
\email{amilcarq@df.ufcg.edu.br}

\affiliation{Departamento de Física, Universidade Federal de Campina Grande Caixa Postal 10071, 58429-900 Campina Grande, Paraíba, Brazil.}


\author{V. B. Bezerra}
\email{valdir@fisica.ufpb.br}
\affiliation{Departamento de Física, Universidade Federal da Paraíba, Caixa Postal 5008, 58051--970, João Pessoa, Paraíba,  Brazil.}

\date{\today}

\begin{abstract}

We investigate the optical properties of a rotating black hole immersed in a Hernquist dark matter halo. The spacetime is obtained from a static Hernquist black hole through the noncomplexification version of the Newman--Janis procedure, leading to a Kerr--like geometry in which the halo contribution is encoded in the radial function $\Delta(r)$, as recently proposed in the literature \cite{AraujoFilho:2026hernquist}.  We first analyze the geodesic structure by deriving the effective potentials, the radial acceleration for null particles, and representative three--dimensional photon trajectories around the event horizon and the ergoregion. In this manner, by using the separability of the Hamilton--Jacobi equation, we obtain the critical impact parameters associated with unstable spherical photon orbits and construct the corresponding shadow contours for a distant observer. We show that the rotation parameter mainly shifts and distorts the shadow, while the Hernquist halo enlarges the photon capture region and increases the apparent shadow size. By comparing the area equivalent shadow diameter with the Event Horizon Telescope measurements of Sgr A$^\ast$ and M87$^\ast$, we obtain upper bounds on the dimensionless halo parameter $\hat{\rho}=M^2\rho$. In particular, Sgr A$^\ast$ gives the strongest restriction, with $\hat{\rho}\sim(2.7-3.8)\times10^{-3}$ at $1\sigma$ and $\hat{\rho}\sim(4.1-5.2)\times10^{-3}$ at $2\sigma$. We also study the strong-- and weak--field gravitational lensing regimes. In the lensing sector, the Hernquist halo affects the strong--field signal by shifting the unstable photon orbit and the critical impact parameter, which control the logarithmic behavior of the deflection angle and the position of the relativistic images. In the weak--field regime, the halo contribution appears already in the leading term of the bending angle, increasing the departure from the Kerr prediction as the density parameter $\rho$ grows. Using the Einstein ring of ESO325-G004, we further constrain the halo parameter as $0\leq \hat{\rho} \lesssim 0.00939$ at $1\sigma$ and $0\leq \hat{\rho} \lesssim 0.01963$ at $2\sigma$.

\end{abstract}


\maketitle

\clearpage

\tableofcontents


\section{Introduction}

The optical appearance of black holes has become one of the most direct ways of testing gravity in the strong--field regime.  Long before horizon-scale observations became possible, it was understood that photons propagating near compact objects may be captured, scattered, or forced to approach unstable null orbits, leaving a characteristic dark region in the observer's sky \cite{Synge:1966okc,Luminet:1979nyg,Bardeen:1973gb,Chandrasekhar:1985kt}.  For rotating black holes, the Kerr geometry provides the standard reference model: the Hamilton--Jacobi equation is separable, the photon region is formed by spherical null orbits, and the shadow boundary is obtained from critical impact parameters associated with unstable photon trajectories \cite{Kerr:1963ud,Carter:1968ks,Bardeen:1973gb,Teo:2003}.  The observation of the horizon scale emission around M87$^\ast$ and Sgr A$^\ast$ by the Event Horizon Telescope (EHT) has turned this theoretical construction into an observational tool, since the angular size and morphology of the bright ring are tightly connected with the photon capture region of the central compact object \cite{EventHorizonTelescope:2019dse,EventHorizonTelescope:2019ths,EventHorizonTelescope:2022wkp,EventHorizonTelescope:2022xqj}.  Although the observed emission ring is not identical to the mathematical shadow boundary, because it depends on the accretion flow, plasma model, optical depth, magnetic field configuration and radiative transfer, its diameter remains a useful estimator of the underlying photon capture scale \cite{Falcke:1999pj,Cunha:2018acu,Gralla:2019xty,Johnson:2019ljv,Perlick:2021aok}.

Black hole shadows are especially useful because they are controlled by a small set of geometrical ingredients.  In the Kerr case, the spin parameter displaces the shadow center and produces the well-known left--right asymmetry caused by frame dragging \cite{Bardeen:1973gb,Hioki:2009na}.  Departures from the Kerr geometry may instead change the size of the photon region, deform the shadow contour, or modify the critical impact parameter that separates captured photons from escaping photons \cite{Johannsen:2010ru,Bambi:2013nla,Atamurotov:2013sca,Cunha:2015yba,Abdujabbarov:2015xqa}.  For this reason, shadow observables have been used to test modified gravity, regular black holes, scalarized compact objects, wormholes and other horizonless alternatives \cite{Berti:2015itd,Cardoso:2019rvt,Psaltis:2020lvx,Vagnozzi:2022moj}.  The same logic applies to black holes embedded in astrophysical environments: if matter surrounding the compact object modifies the effective gravitational field felt by photons, then the shadow records this modification through the photon sphere, photon shell and capture cross section \cite{Perlick:2015vta,Konoplya:2019sns,Jusufi:2019nrn}.

Dark matter halos (DMH) constitute the natural galactic environment in which supermassive black holes are expected to reside. Their existence is supported by several independent observations, including galactic rotation curves, stellar and gas dynamics, gravitational lensing, the formation of large-scale structure, and the cosmic microwave background \cite{Rubin:1980zd,Begeman:1991iy,Bertone:2004pz,Freese:2008cz,Planck:2018vyg,Wechsler:2018pic}. At the phenomenological level, these effects are commonly incorporated through effective density profiles. The Navarro--Frenk--White profile is widely used in cold dark matter simulations, while Einasto-- and Burkert--type profiles, as well as Dehnen--like families, provide complementary descriptions of cusped or cored galactic distributions \cite{Navarro:1996gj,Dutton:2014xda,Burkert:1995yz,Dehnen:1993}. In this context, the Hernquist profile is particularly useful because it furnishes an analytic potential-density model for spherical galaxies and bulges, possesses a finite total mass, and displays an outer falloff compatible with compact stellar systems \cite{Hernquist:1990be}. In other words, placing a black hole inside a Hernquist halo is not merely a formal deformation of the Kerr or Schwarzschild geometries; rather, it provides a controlled way to relate strong-field observables to the matter distribution of the host galaxy. This idea has motivated a growing literature on black holes surrounded by DMH, where the halo contribution modifies horizon structure, geodesic motion, photon regions, accretion properties, shadows, quasinormal modes, superradiant behavior, thermodynamics, and gravitational wave fluxes \cite{Sadeghian:2013laa,Cardoso:2021wlq,Figueiredo:2023gas,Anjum:2023dark,Yang:2023hci,Belchior:2026vwf,AlBadawi:2024xxx,Jha:2025hernquist,Lobo:2025m60shadow,Lobo:2025m60structure,Liu:2022qnm,Nieto:2025hernquist,Al-Badawi:2026kkw,Ahmed:2026gvw,Al-Badawi:2025ipr,K1,K2,K3,K4}.

The optical sector is especially sensitive to this environmental correction. The shadow boundary is determined by unstable null geodesics, so any halo--induced modification of the radial metric function changes the effective photon potential, the location of spherical photon orbits, and the critical impact parameters that separate captured from scattered light rays \cite{Konoplya:2019sns,Jusufi:2019nrn,Anjum:2023dark,Figueiredo:2023gas}. For Hernquist--type backgrounds, recent analyses have shown that the halo parameters can shift the photon sphere, modify the shadow size, affect the observed intensity profile for different accretion prescriptions, and leave measurable imprints on weak-lensing, accretion--disk observables \cite{Jha:2025hernquist,Shi:2025hernquist,Nieto:2025hernquist,HeydariFard:2026hernquist} and other asptects \cite{Heidari:2026swh,Heidari:2026eim}. Nevertheless, most studies have either considered static configurations or focused on halo profiles different from the Hernquist distribution. A rotating Hernquist black hole therefore deserves a separate optical analysis. Indeed, the spin and the halo parameter affect the image in physically different ways: rotation mainly displaces the shadow center and introduces left--right asymmetry through frame dragging, whereas the Hernquist contribution changes the radial capture scale of photons. The simultaneous study of null geodesics, shadows, and gravitational lensing is then a direct way to disentangle these two effects and to test whether the surrounding halo can be constrained by horizon--scale and weak--field observations.

Gravitational lensing provides an independent probe of the same photon sector.  In the weak--field regime, light rays pass far from the lens and the deflection angle can be expanded in powers of the inverse impact parameter, leading to the standard post--Newtonian description of lensing by stars, galaxies and clusters \cite{Weinberg:1972kfs,Schneider:1992,Bartelmann:2010fz}.  In the strong--field regime, photons approach the unstable photon orbit, the deflection angle develops a logarithmic divergence, and an infinite sequence of relativistic images can be formed near the optical axis \cite{Darwin:1959,Virbhadra:1999nm,Bozza:2002zj,Perlick:2003vg,Tsukamoto:2016jzh}.  The limiting angular position of these images, their angular separation, the magnification ratio and the time delay between different winding trajectories are governed by the critical impact parameter and the strong-deflection coefficients \cite{Bozza:2003cp,Bozza:2008ev,Eiroa:2002mk}.  Therefore, shadows and lensing are complementary rather than redundant: both are controlled by unstable null geodesics, but they translate the photon region structure into different observational quantities.

The weak--lensing sector is also sensitive to DMH.  A halo contribution modifies the bending angle at large impact parameter and may affect Einstein-ring sizes in galaxy-scale lensing systems \cite{Keeton:2005jd,Bartelmann:2010fz}.  For rotating black holes, the spin contribution enters the deflection angle together with the mass terms, while environmental corrections may appear already in the leading or next-to-leading orders, depending on the asymptotic behavior of the metric functions \cite{Ono:2017pie,Ishihara:2016vdc,Ovgun:2019wej,Vachher:2025jsq,Li:2021lhr,He:2022yco,Kumar:2025bim}.  This makes weak lensing a useful counterpart to EHT shadow bounds: the shadow probes the photon capture region near the black hole, while the weak deflection angle probes the cumulative effect of the geometry along trajectories with larger impact parameters.

Recently, a rotating black hole immersed in a Hernquist DMH was constructed from a static Hernquist black hole \cite{AraujoFilho:2026hernquist} through the noncomplexification version of the Newman--Janis procedure \cite{Newman:1965tw,AzregAinou:2014pra}. 
The resulting spacetime has a Kerr--like angular sector, whereas the halo contribution is encoded in the radial function $\Delta(r)$. 
The Kerr solution is recovered when the halo density parameter vanishes, and the static Hernquist black hole is recovered when the spin is switched off. This structure is well suited for optical studies because the separability properties of the Kerr-like sector allow one to derive the null geodesic equations and the critical impact parameters, while the Hernquist contribution modifies the radial potential that determines photon capture, lensing and the shadow scale.

In the present work, we investigate the optical traces of this rotating Hernquist black hole.  First, we derive the geodesic equations and analyze the effective potentials governing particle motion.  For null trajectories, we study the radial acceleration and display representative three-dimensional photon trajectories around the event horizon and the ergoregion.  Then, using the separability of the Hamilton--Jacobi equation, we determine the critical impact parameters of unstable spherical photon orbits and construct the corresponding shadow contours for a distant observer.  We show that the spin parameter mainly shifts and distorts the shadow, whereas the Hernquist halo enlarges the photon capture region and increases the apparent shadow size.  By comparing the area-equivalent shadow diameter with the EHT measurements of Sgr A$^\ast$ and M87$^\ast$, we obtain upper bounds on the dimensionless halo parameter $\hat{\rho}=M^2\rho$.

We also study gravitational lensing in both the strong-- and weak--field regimes.  In the strong-deflection limit, we compute the angular position of the relativistic images, their separation, the relative magnification and the time delay for Sgr A$^\ast$ and M87$^\ast$.  In the weak--field regime, we derive the bending angle as an expansion in the inverse impact parameter and identify the Hernquist corrections to the Kerr result.  Finally, using the Einstein ring of ESO325-G004 \cite{Smith:2005ru,Smith:2013mda}, we constrain the halo parameter independently from weak--lensing data.  This combination of shadow observables, relativistic images and Einstein--ring bounds provides a unified optical test of rotating black holes immersed in Hernquist DMHs.


\section{The rotating black hole solution}
\label{sec:rotating_solution}

Recently, a rotating black hole surrounded by a Hernquist DMH was proposed in Ref. \cite{AraujoFilho:2026hernquist}. Since the present work is devoted to the optical sector of this geometry, namely, geodesics, shadows and gravitational lensing, we shall not repeat the complete derivation of the rotating spacetime. Instead, we briefly collect the geometrical ingredients that will be used throughout the following sections.

The starting point is the static and spherically symmetric line element
\begin{equation}
\mathrm{d}s^{2} = -f(r)\,\mathrm{d}t^{2} + \frac{\mathrm{d}r^{2}}{f(r)} + r^{2}\mathrm{d}\Omega^{2},
\label{static_hernquist_metric}
\end{equation}
where
\begin{equation}
    f(r)
    =
    1-\frac{2M}{r}
    -
    \frac{4\pi \rho r_{s}^{3}}{r+r_{s}} .
    \label{hernquist_seed_function}
\end{equation}
Here, $M$ represents the black hole mass, whereas $\rho$ and $r_s$ characterize, respectively, the density scale and the length scale associated with the surrounding Hernquist distribution. In the absence of the halo contribution, $\rho \rightarrow 0$, the Schwarzschild lapse function is recovered.

The rotating counterpart is obtained by applying the noncomplexification version of the Newman--Janis procedure. In this prescription, the rotation parameter $a$ is introduced while the radial functions of the seed metric are promoted to real functions of $r$, $a$ and $\theta$. The resulting geometry can be written in Boyer--Lindquist--type coordinates as
\begin{equation}
\begin{split}
    \mathrm{d}s^{2} ={}&
    -\left(
    \frac{\Delta(r)-a^{2}\sin^{2}\theta}{\Sigma}
    \right)\mathrm{d}t^{2}
    +
    \frac{\Sigma}{\Delta(r)}\,\mathrm{d}r^{2}
    +
    \Sigma\,\mathrm{d}\theta^{2}
    \\[0.2cm]
    &-2a\sin^{2}\theta
    \left[
    1-
    \frac{\Delta(r)-a^{2}\sin^{2}\theta}{\Sigma}
    \right]\mathrm{d}t\,\mathrm{d}\varphi
    \\[0.2cm]
    &+
    \frac{\sin^{2}\theta}{\Sigma}
    \left[
    \left(r^{2}+a^{2}\right)^{2}
    -
    a^{2}\Delta(r)\sin^{2}\theta
    \right]\mathrm{d}\varphi^{2},
\end{split}
\label{rotating_hernquist_metric}
\end{equation}
with $\Sigma = r^{2}+a^{2}\cos^{2}\theta$ and $\Delta(r)=r^{2}f(r)+a^{2}$. Using Eq. \eqref{hernquist_seed_function}, the radial function assumes the explicit form
\begin{equation}
    \Delta(r)
    =
    r^{2}-2Mr+a^{2}
    -
    \frac{4\pi \rho r_{s}^{3}r^{2}}{r+r_{s}} .
    \label{delta_general_hernquist}
\end{equation}

The metric \eqref{rotating_hernquist_metric} has the same Kerr--like structure in its angular sector, but the function $\Delta(r)$ carries the information about the DMH. Therefore, the standard Kerr spacetime is obtained in the limit $\rho\rightarrow 0$, while the static Hernquist black hole is recovered when $a\rightarrow 0$. If both limits are taken simultaneously, one returns to the Schwarzschild geometry.

The spacetime is stationary and axisymmetric, so that $\partial_t$ and $\partial_\varphi$ are Killing vectors. This property guarantees the conservation of the particle energy and azimuthal angular momentum along geodesic motion, which will be used in the next section.

The possible horizons are determined by $\Delta(r)=0$. In the parametrization $r_s=2M$, this condition may be written as the cubic equation $\left(r^{2}-2Mr+a^{2}\right)(r+2M)- 32\pi \rho M^{3}r^{2} = 0.$. The event horizon $r_h$ corresponds to the largest positive root. On the other hand, the stationary limit surface follows from $g_{tt}=0$, namely, $\Delta(r)-a^{2}\sin^{2}\theta=0$. The region between the outer stationary limit surface and the event horizon defines the ergoregion.


\section{Geodesics}

In this section, we investigate the motion of test particles in the rotating black hole spacetime surrounded by a Hernquist DMH. Since the geometry is stationary and axisymmetric, the vectors $\partial_t$ and $\partial_\phi$ are Killing vectors. In this manner, the energy and the azimuthal angular momentum of a test particle are conserved along the geodesic flow. The influence of the halo enters the geodesic equations through the radial function $\Delta(r)$, while the rotation parameter $a$ controls the coupling between the temporal and azimuthal directions.

We start from the point particle Lagrangian
\ie
\mathcal{L}=g_{\mu\nu}\dot{x}^{\mu}\dot{x}^{\nu},
\fe
where the dot denotes differentiation with respect to the affine parameter $\lambda$. The normalization condition is chosen as $\mathcal{L}=-1,0,+1$ for timelike, null and spacelike geodesics, respectively. For the metric under consideration, we obtain
\ie
\begin{split}
\mathcal{L}  = & -\left[\frac{\Delta(r)-a^2\sin^2\theta}{\Sigma}\right]\dot{t}^{\,2} +\frac{\Sigma}{\Delta(r)}\dot{r}^{\,2} +\Sigma \dot{\theta}^{\,2}  \\ & -2a\sin^2\theta \left[1-\frac{\Delta(r)-a^2\sin^2\theta}{\Sigma}\right]\dot{t}\dot{\phi} +\frac{\sin^2\theta}{\Sigma} \left[(r^2+a^2)^2-a^2\Delta(r)\sin^2\theta\right]\dot{\phi}^{\,2}.
\label{lagrangian_full_geodesics}
\end{split}
\fe
In the following, we first restrict the motion to the equatorial plane, $\theta=\pi/2$ and $\dot{\theta}=0$. In this plane, $\Sigma=r^2$, and the Lagrangian becomes
\ie
\begin{split}
\mathcal{L}  = &
-\left[\frac{\Delta(r)-a^2}{r^2}\right]\dot{t}^{\,2}
+\frac{r^2}{\Delta(r)}\dot{r}^{\,2}  \\
& -2a\left[1-\frac{\Delta(r)-a^2}{r^2}\right]\dot{t}\dot{\phi}
+\frac{1}{r^2}\left[(r^2+a^2)^2-a^2\Delta(r)\right]\dot{\phi}^{\,2}.
\label{lagrangian_equatorial_geodesics}
\end{split}
\fe

It is useful to introduce the functions
\ie
A(r)=\frac{\Delta(r)-a^2}{r^2}, \qquad
B(r)=a\left[1-\frac{\Delta(r)-a^2}{r^2}\right],
\fe
and
\ie
C(r)=\frac{(r^2+a^2)^2-a^2\Delta(r)}{r^2}.
\fe
In terms of these quantities, Eq.~\eqref{lagrangian_equatorial_geodesics} can be written as
\ie
\mathcal{L}=-A\dot{t}^{\,2}-2B\dot{t}\dot{\phi}
+C\dot{\phi}^{\,2}+\frac{r^2}{\Delta(r)}\dot{r}^{\,2}.
\fe
The conserved energy $E$ and angular momentum $L$ are then given by
\ie
E=-p_t=A\dot{t}+B\dot{\phi},
\label{energy_equatorial_geodesics}
\fe
and
\ie
L=p_\phi=-B\dot{t}+C\dot{\phi}.
\label{angular_equatorial_geodesics}
\fe
Solving Eqs.~\eqref{energy_equatorial_geodesics} and
\eqref{angular_equatorial_geodesics} for $\dot{t}$ and $\dot{\phi}$, and using the identity
\ie
A(r)C(r)+B^2(r)=\Delta(r),
\fe
we find
\ie
\dot{t}=\frac{C(r)E-B(r)L}{\Delta(r)},
\label{tdot_equatorial_geodesics}
\fe
and
\ie
\dot{\phi}=\frac{B(r)E+A(r)L}{\Delta(r)}.
\label{phidot_equatorial_geodesics}
\fe
Therefore, the equatorial radial equation assumes the compact form
\ie
\dot{r}^{\,2} = \frac{1}{r^2} \left[C(r)E^2-2B(r)EL-A(r)L^2+\mathcal{L}\Delta(r) \right].
\label{radial_equation_compact}
\fe

The previous equation can be factorized in terms of two effective branches,
\ie
\dot{r}^{\,2} = \frac{C(r)}{r^2} \left[E-\mathcal{V}_{+}(r)\right] \left[E-\mathcal{V}_{-}(r)\right],
\label{radial_equation_potentials}
\fe
where
\ie
\mathcal{V}_{\pm}(r) = \frac{ B(r)L \pm \sqrt{ B^2(r)L^2+C(r)\left[A(r)L^2-\mathcal{L}\Delta(r)\right] } }{C(r)}.
\label{effective_potentials_general}
\fe
The condition $\dot{r}^{\,2}\geq0$ selects the regions where the motion is allowed. Outside the outer horizon, $C(r)$ is positive for the parameter range considered here, so the allowed domains are determined by the sign of $\left[E-\mathcal{V}_{+}(r)\right]\left[E-\mathcal{V}_{-}(r)\right]$. In particular, the interval between the two branches corresponds to a forbidden region for fixed energy, whereas the turning points of the trajectory are obtained from $E=\mathcal{V}_{+}$ or $E=\mathcal{V}_{-}$.

\begin{figure}
    \centering
    \includegraphics[scale=0.42]{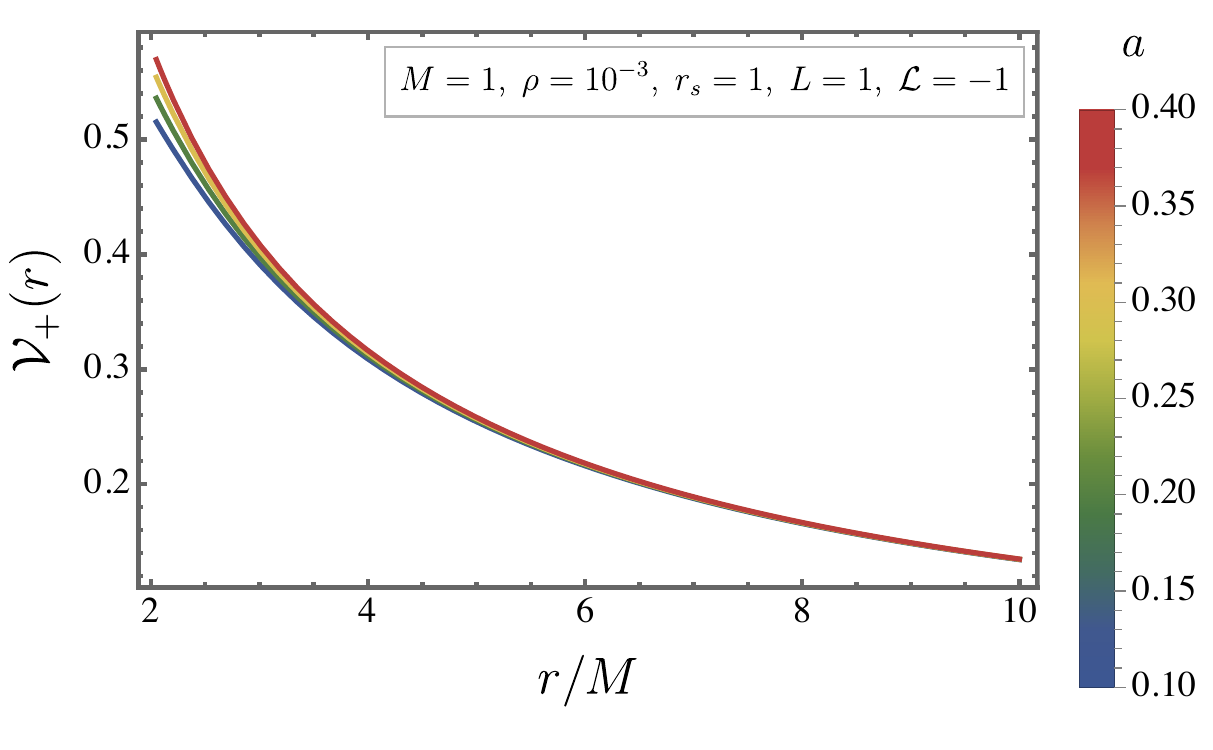}
    \includegraphics[scale=0.42]{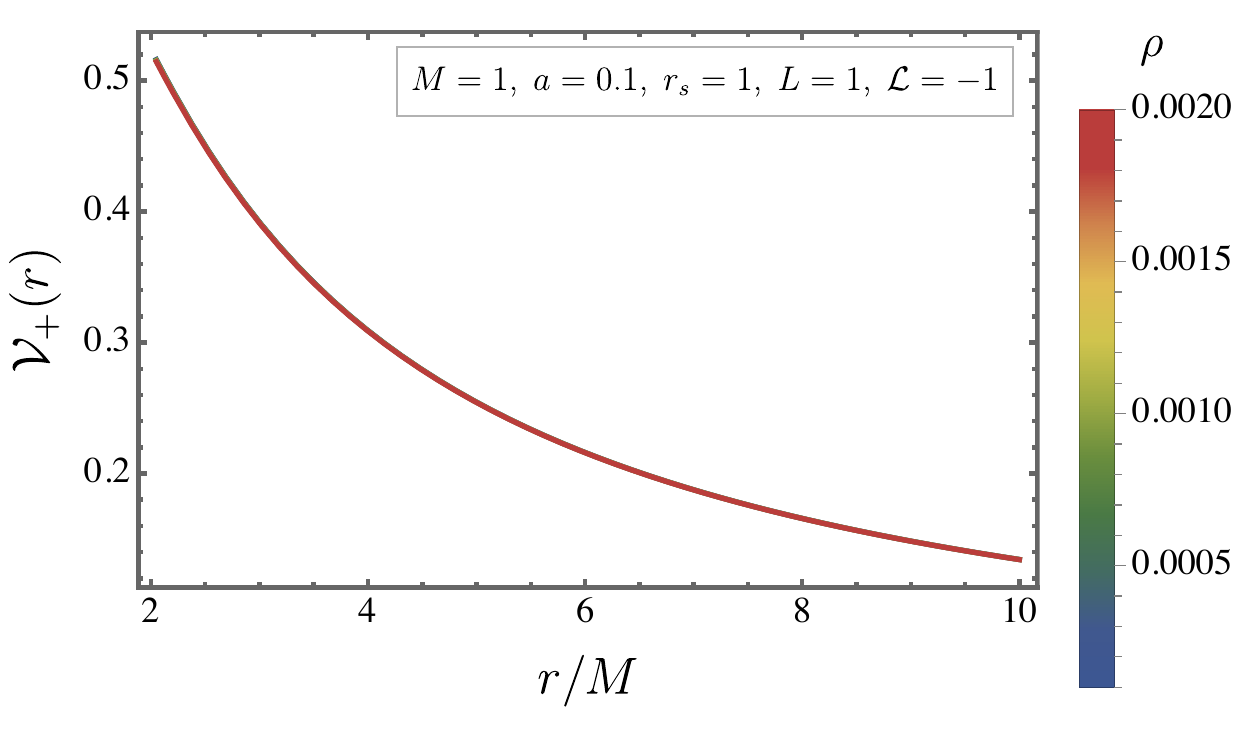}
    \caption{Effective potential $\mathcal{V}_{+}$ for timelike equatorial geodesics. The left panel shows the effect of varying the rotation parameter $a$ at fixed $\rho$, while the right panel displays the effect of varying the Hernquist density parameter $\rho$ at fixed $a$. The spin produces a stronger deformation in the near-horizon region, whereas the halo contribution changes the radial profile through $\Delta(r)$.}
    \label{potentialsplus}
\end{figure}

\begin{figure}
    \centering
    \includegraphics[scale=0.42]{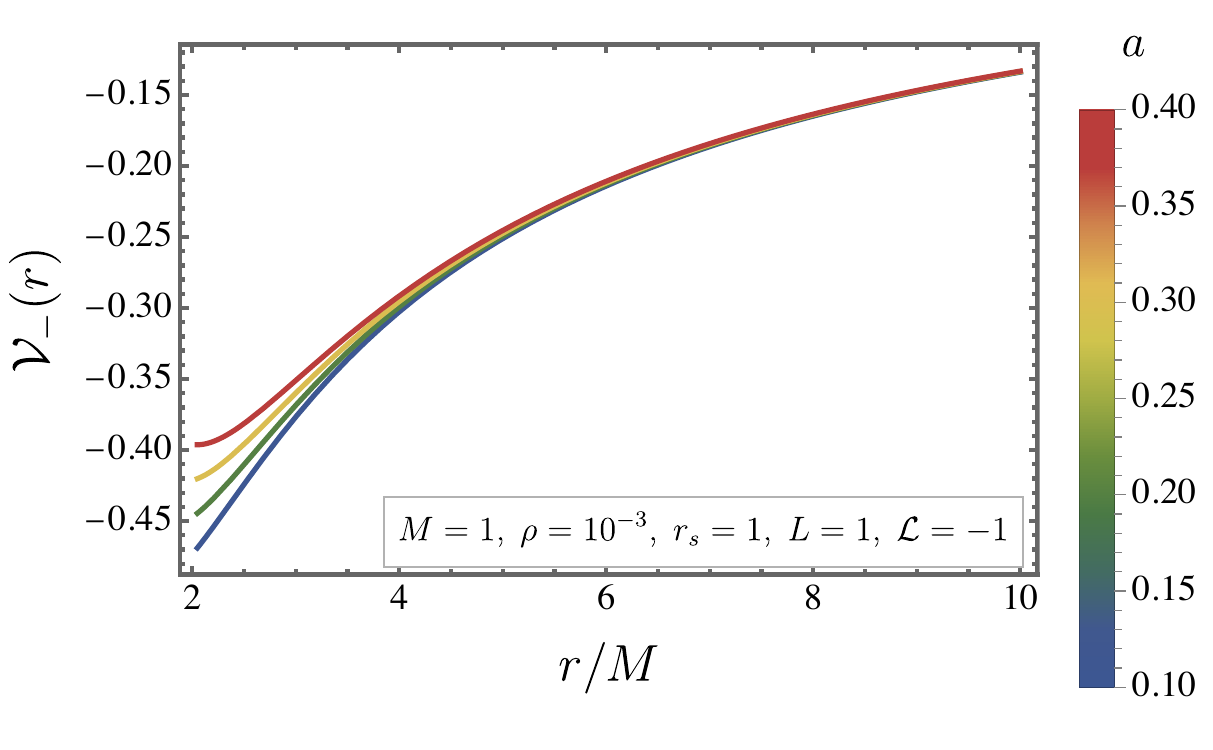}
    \includegraphics[scale=0.42]{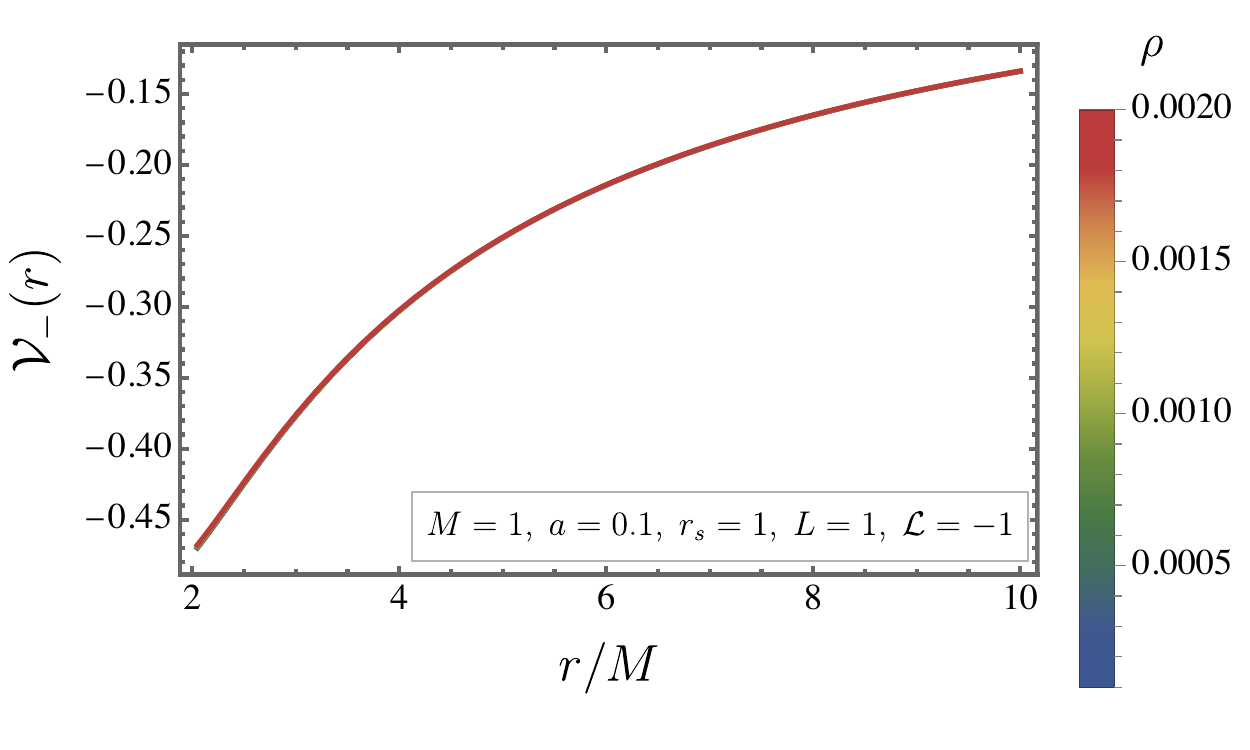}
    \caption{Effective potential $\mathcal{V}_{-}$ for timelike equatorial geodesics. The two panels use the same parameter choices as in Fig.~\ref{potentialsplus}. The left branch approaches zero from below as $r$ increases, while the strongest deviations occur close to the compact object, where the frame--dragging and halo corrections are more relevant.}
    \label{potentialsminus}
\end{figure}

Figures~\ref{potentialsplus} and \ref{potentialsminus} show the behavior of the two branches for timelike motion, $\mathcal{L}=-1$. In Fig.~\ref{potentialsplus}, the upper branch $\mathcal{V}_{+}$ is displayed as a function of $r/M$. The left panel isolates the role of the rotation parameter $a$, while the right panel shows the dependence on the Hernquist density parameter $\rho$. The spin dependence is more visible in the strong--field region, where the frame--dragging contribution modifies the energy required for a particle to reach a given radial turning point. By contrast, the variation with $\rho$ is milder for the small values used in the plot, although the halo contribution still changes the radial capture scale through $\Delta(r)$.

The lower branch $\mathcal{V}_{-}$, shown in Fig.~\ref{potentialsminus}, complements this picture. Together, $\mathcal{V}_{+}$ and $\mathcal{V}_{-}$ determine the energy window that cannot be crossed by a particle with fixed $L$. The separation between the branches is mainly affected by the rotation near the black hole, while the halo parameter produces a radial deformation of the potential. This behavior is consistent with the optical analysis performed later: the parameter $a$ is responsible for the left--right asymmetry and displacement of the shadow, whereas $\rho$ predominantly changes the size of the photon capture region.

\subsection{Radial acceleration of null geodesics}

We now specialize the previous discussion to null geodesics, for which $\mathcal{L}=0$. In this case, Eq.~\eqref{radial_equation_compact} reduces to
\ie
\dot{r}^{\,2} = \frac{1}{r^2} \left[ C(r)E^2-2B(r)EL-A(r)L^2 \right].
\label{null_radial_equation_compact}
\fe
Equivalently, using the explicit form of $A(r)$, $B(r)$ and $C(r)$, this equation may be written as
\ie
\dot{r}^{\,2} = \frac{1}{r^4} \left\{ \left[E(r^2+a^2)-aL\right]^2 -\Delta(r)\left(aE-L\right)^2 \right\}.
\label{null_radial_equation_explicit}
\fe
This form makes clear that the halo changes the photon motion only through the radial function $\Delta(r)$, whereas the terms proportional to $a$ encode the rotational dragging of null rays.

Let
\ie
F(r)\equiv \frac{C(r)}{r^2} \left[E-\mathcal{V}_{+}(r)\right] \left[E-\mathcal{V}_{-}(r)\right],
\fe
so that $\dot{r}^{\,2}=F(r)$. Differentiating with respect to the affine parameter gives
\ie
2\dot{r}\ddot{r}=F'(r)\dot{r},
\fe
and therefore
\ie
\begin{split}
\ddot{r} =& \frac{1}{2} \left(\frac{C(r)}{r^2}\right)' \left[E-\mathcal{V}_{+}(r)\right] \left[E-\mathcal{V}_{-}(r)\right] \\ &- \frac{C(r)}{2r^2} \left[ \mathcal{V}_{+}'(r)\left(E-\mathcal{V}_{-}(r)\right) + \mathcal{V}_{-}'(r)\left(E-\mathcal{V}_{+}(r)\right) \right].
\label{radial_acceleration_general}
\end{split}
\fe
At a radial turning point, the radial velocity vanishes. If the turning point is located on the upper branch, $E=\mathcal{V}_{+}$, one obtains
\ie
\ddot{r}_{+} = -\frac{C(r)}{2r^2} \mathcal{V}_{+}'(r) \left[ \mathcal{V}_{+}(r)-\mathcal{V}_{-}(r) \right].
\label{radial_acceleration_plus}
\fe
Similarly, for a turning point on the lower branch, $E=\mathcal{V}_{-}$, we have
\ie
\ddot{r}_{-} = -\frac{C(r)}{2r^2} \mathcal{V}_{-}'(r) \left[ \mathcal{V}_{-}(r)-\mathcal{V}_{+}(r) \right].
\label{radial_acceleration_minus}
\fe

\begin{figure}
    \centering
    \includegraphics[scale=0.42]{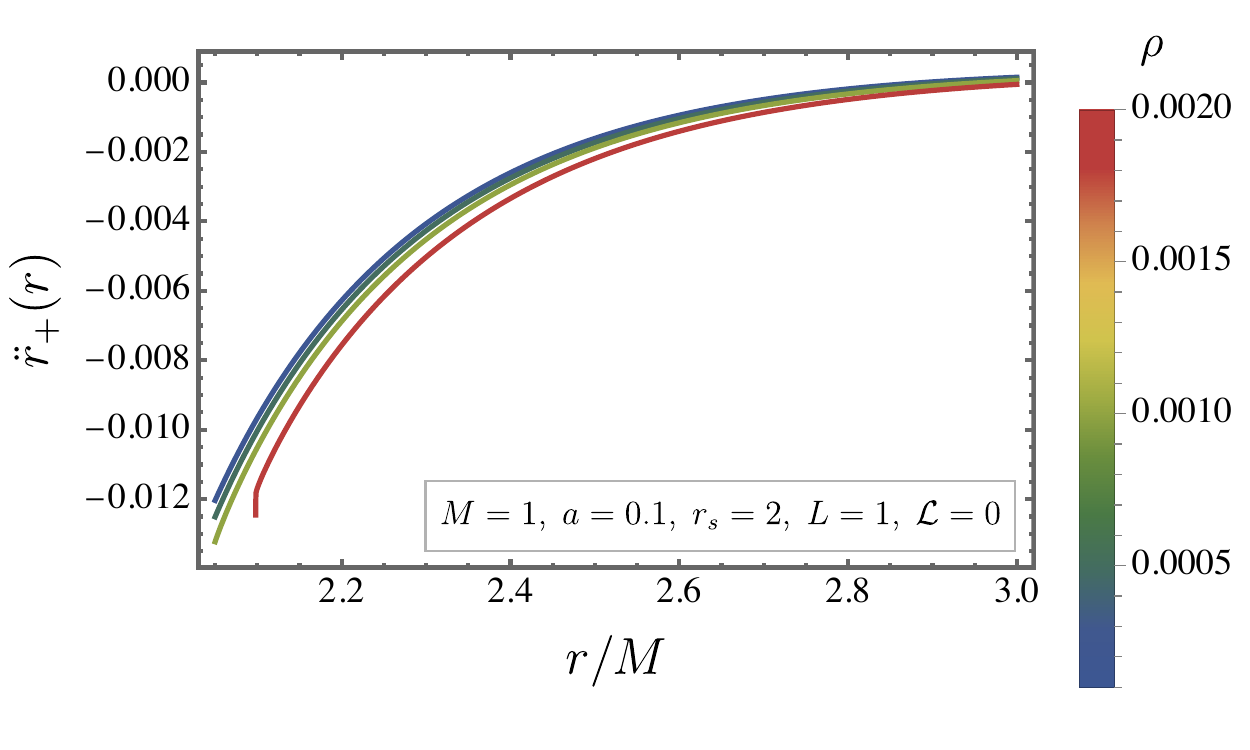}
    \includegraphics[scale=0.42]{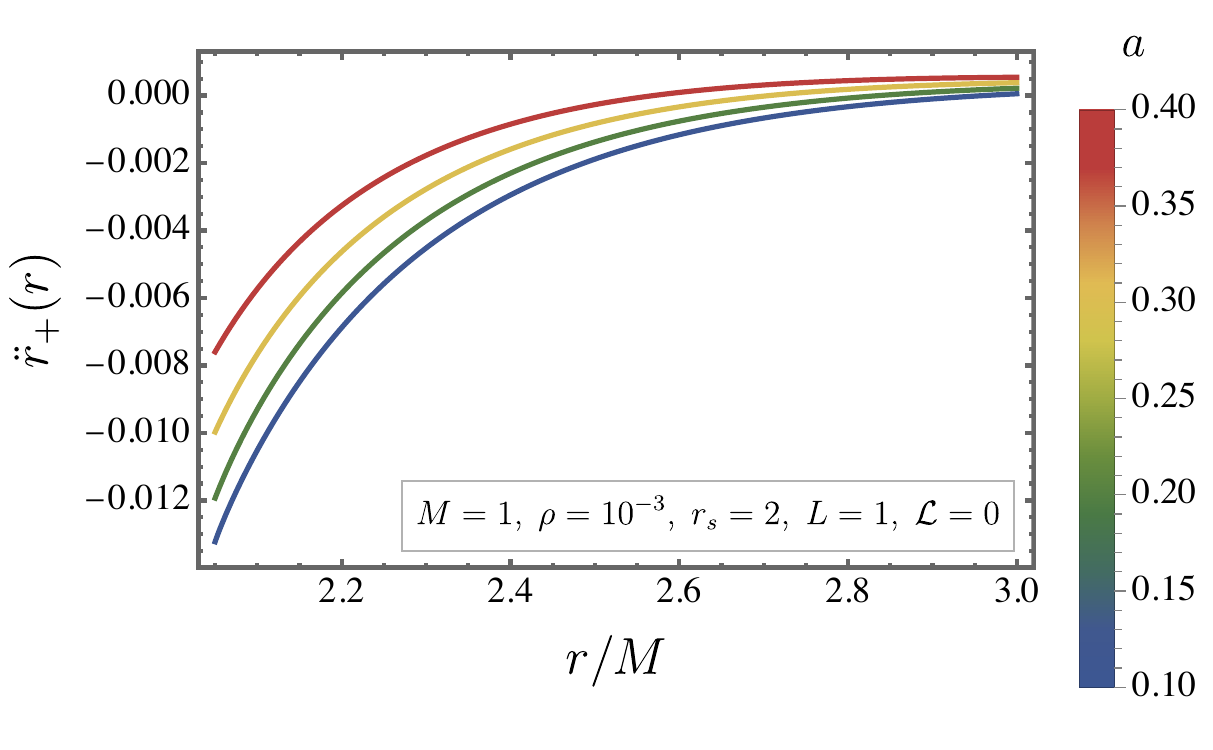}
    \caption{Radial acceleration $\ddot{r}_{+}$ for null geodesics evaluated at the upper turning branch. The left panel shows the dependence on the Hernquist density parameter $\rho$ at fixed $a$, while the right panel shows the dependence on the rotation parameter $a$ at fixed $\rho$. In the displayed range, $\ddot{r}_{+}<0$, so the corresponding turning branch is associated with inward radial acceleration.}
    \label{radialaccel}
\end{figure}

Figure~\ref{radialaccel} displays $\ddot{r}_{+}$ for null geodesics. The acceleration is negative in the plotted interval, indicating that a photon placed at the corresponding turning branch is driven towards smaller radii. As $r$ increases, the magnitude of $\ddot{r}_{+}$ decreases and the curve tends to zero, as expected outside the strong-field region. The left panel shows that increasing $\rho$ slightly strengthens the inward radial pull near the compact object, since the Hernquist halo changes the effective radial function $\Delta(r)$. The right panel shows that the rotation parameter produces a more pronounced modification in the same region.

\begin{figure}
    \centering
    \includegraphics[scale=0.35]{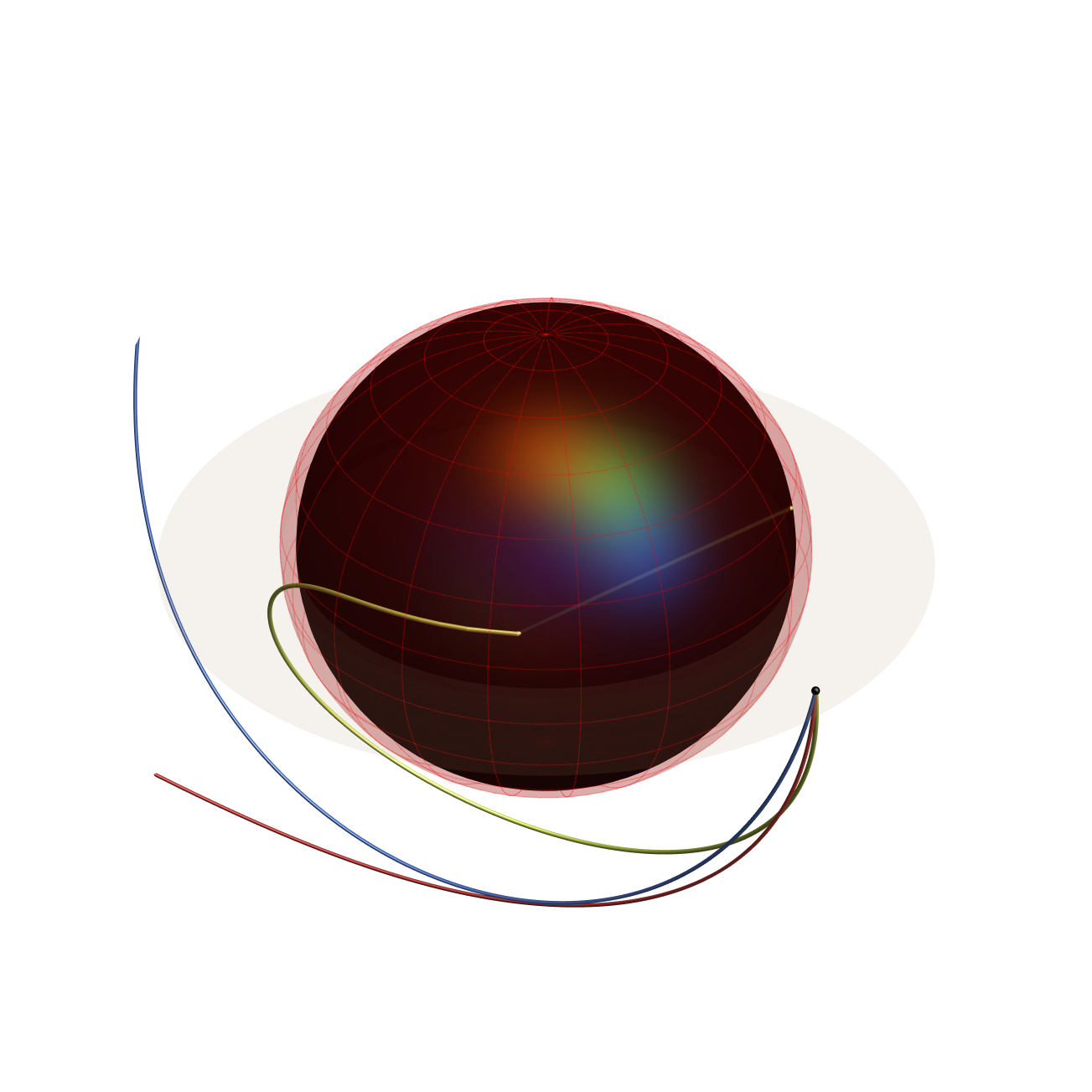}
    \includegraphics[scale=0.35]{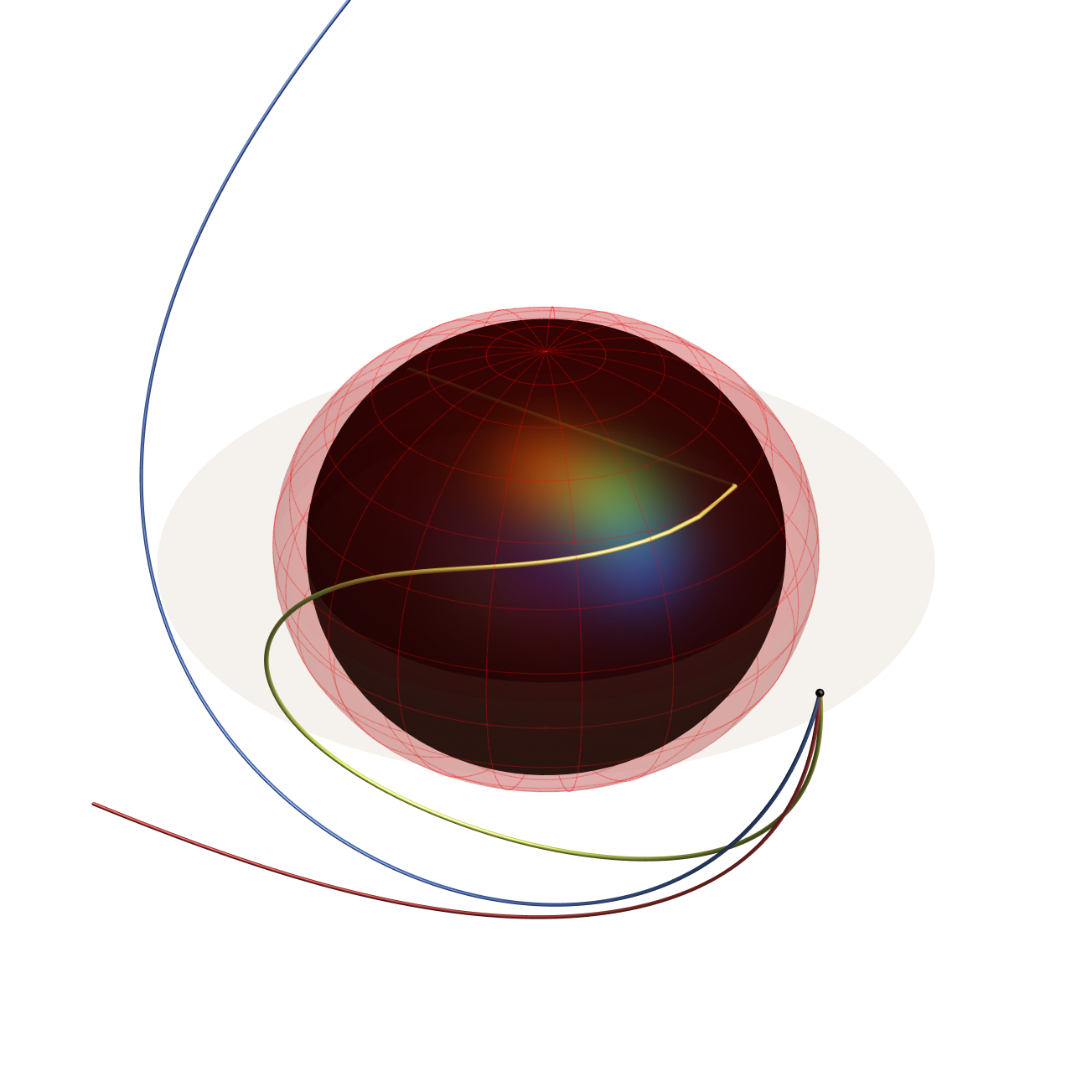}
    \includegraphics[scale=0.35]{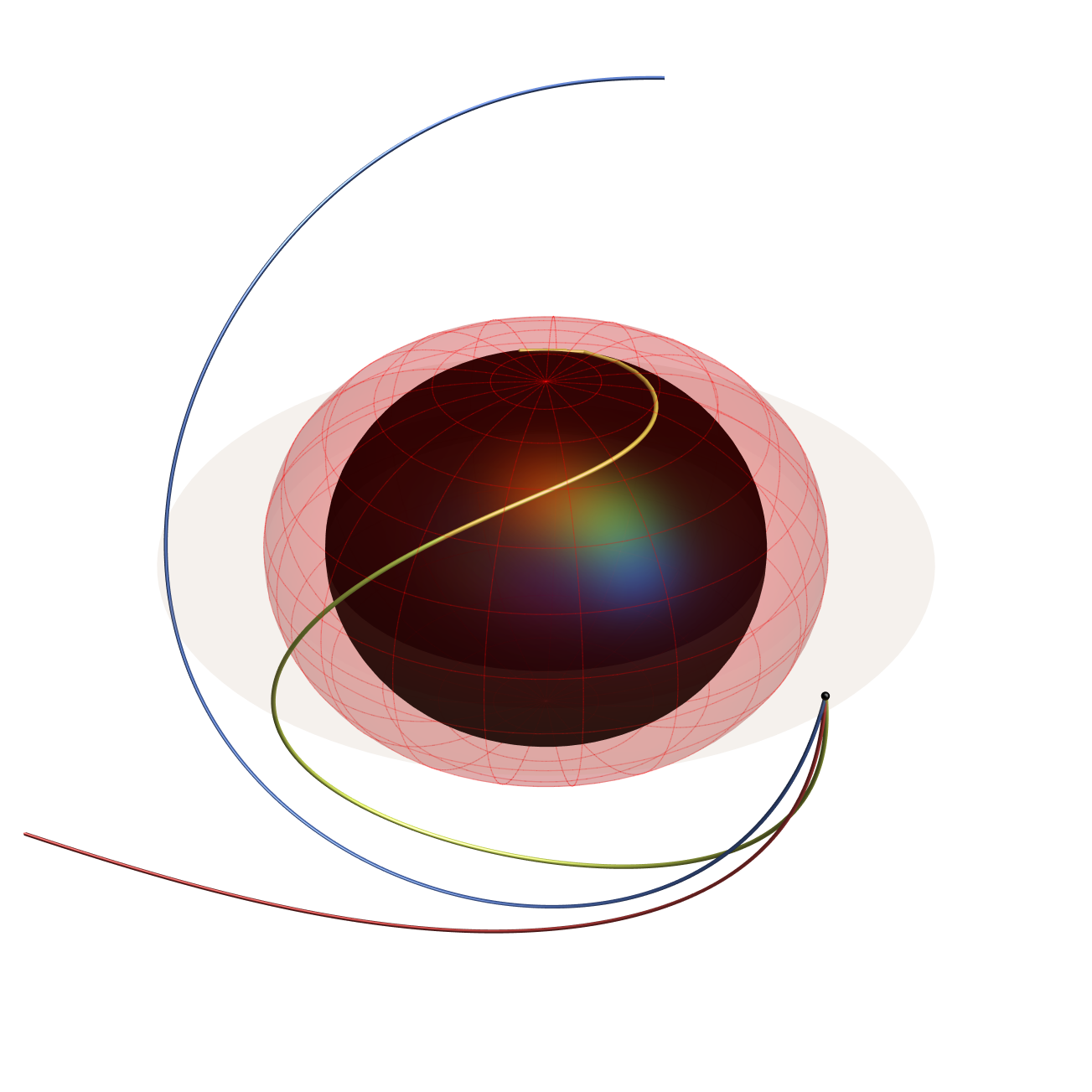}
    \includegraphics[scale=0.35]{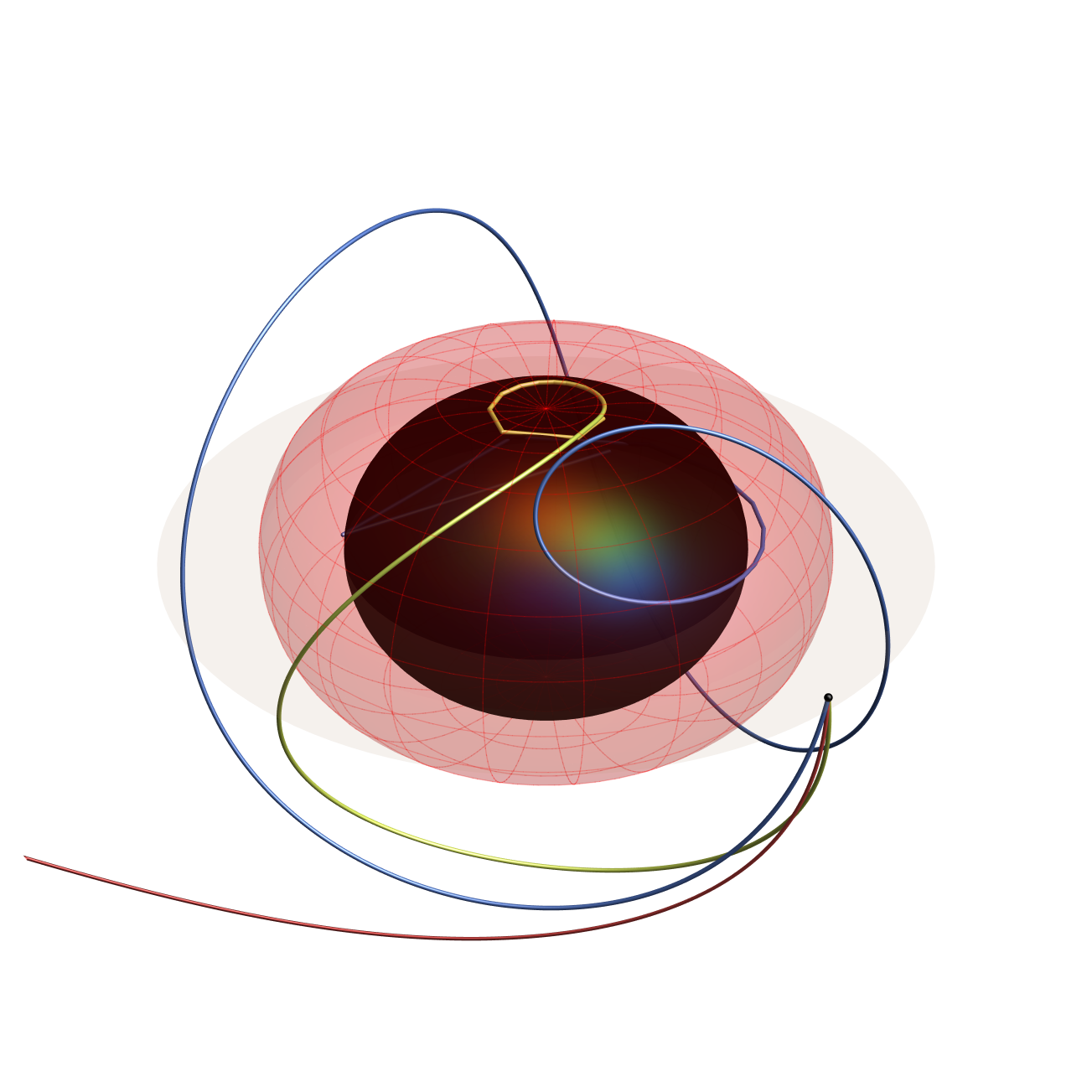}
    \caption{Representative three--dimensional geodesic trajectories around the rotating black hole immersed in a Hernquist DMH. The black surface represents the event horizon, whereas the red surface represents the stationary limit surface. The region between these two surfaces corresponds to the ergoregion. The different colored curves were obtained from distinct initial conditions and illustrate captured, scattered and strongly deflected trajectories.}
    \label{geodesics3}
\end{figure}

Finally, Fig.~\ref{geodesics3} illustrates representative three--dimensional geodesic trajectories around the rotating Hernquist black hole. The black surface denotes the event horizon, while the red surface represents the stationary limit surface. The region between them corresponds to the ergoregion. The curves show that photons passing sufficiently close to the compact object are strongly affected by the rotational dragging and by the deformation of the radial potential. Depending on the initial conditions, a trajectory may be captured by the black hole, scatter back to large distances, or execute a winding motion near the ergoregion before escaping. The role of the halo is not to introduce an additional azimuthal asymmetry, but to alter the radial scale at which capture and strong deflection occur. This point is important for the shadow and lensing analyses below, where the parameter $\rho$ mainly changes the size of the photon capture region, while $a$ shifts and distorts the observed contour.

\section{Photon region, shadows, and observational constraints}

\subsection{Spherical photon orbits and celestial coordinates}

We now investigate the shadow cast by the rotating black hole immersed in a Hernquist DMH. The boundary of the shadow is determined by the unstable photon region, since photons asymptotically approaching these orbits separate the trajectories captured by the black hole from those escaping to spatial infinity. In the present spacetime, the separability of the null Hamilton--Jacobi equation is preserved by the Kerr--like structure of the angular sector, while the halo contribution is encoded in the radial function $\Delta(r)$.

For a massless particle, the Hamilton--Jacobi equation reads
\ie
g^{\mu\nu}\frac{\partial S}{\partial x^{\mu}}
\frac{\partial S}{\partial x^{\nu}}=0,
\fe
where we use the separable ansatz
\ie
S=-Et+L_z\phi+S_r(r)+S_{\theta}(\theta).
\fe
Here $E=-p_t$ and $L_z=p_{\phi}$ are the conserved energy and azimuthal angular momentum associated with the Killing vectors $\partial_t$ and $\partial_{\phi}$, respectively. In addition, the separability of the system introduces the Carter constant $\mathcal{Q}$. The corresponding first--order equations for null geodesics are
\ie
\begin{split}
\Sigma \dot{t} &=
\frac{r^{2}+a^{2}}{\Delta(r)}
\left[E(r^{2}+a^{2})-aL_z\right]
-a\left(aE\sin^{2}\theta-L_z\right),\\
\Sigma \dot{\phi} &=
\frac{a}{\Delta(r)}
\left[E(r^{2}+a^{2})-aL_z\right]
-\left(aE-\frac{L_z}{\sin^{2}\theta}\right),\\
\Sigma^{2}\dot{r}^{2} &=
\left[E(r^{2}+a^{2})-aL_z\right]^2
-\Delta(r)\left[\mathcal{Q}+(aE-L_z)^2\right]
\equiv \mathcal{R}(r),\\
\Sigma^{2}\dot{\theta}^{2} &=
\mathcal{Q}
-\left(\frac{L_z^{2}}{\sin^{2}\theta}-a^{2}E^{2}\right)\cos^{2}\theta
\equiv \Theta(\theta).
\end{split}
\fe
For later convenience, we introduce the dimensionless impact parameters
\ie
\xi=\frac{L_z}{E},\qquad
\eta=\frac{\mathcal{Q}}{E^2}.
\fe
After dividing the radial and angular potentials by $E^2$, one obtains
\ie
\begin{split}
\frac{\mathcal{R}(r)}{E^2}
&=\left[(r^{2}+a^{2})-a\xi\right]^2
-\Delta(r)\left[\eta+(a-\xi)^2\right],\\
\frac{\Theta(\theta)}{E^2}
&=\eta-\left(\frac{\xi^2}{\sin^2\theta}-a^2\right)\cos^2\theta .
\end{split}
\fe

The shadow edge is generated by unstable spherical photon orbits, namely null geodesics with constant radius $r=r_p$ outside the event horizon. These orbits are selected by
\ie
\mathcal{R}(r_p)=0,\qquad
\left.\frac{\mathrm{d}\mathcal{R}}{\mathrm{d}r}\right|_{r=r_p}=0,
\qquad r_p>r_h .
\fe
The unstable branch is identified by the sign of the second radial derivative, with the photon shell providing the set of critical trajectories that form the apparent boundary of the black hole. Solving the two previous equations for $\xi$ and $\eta$ gives the critical impact parameters
\ie
\begin{split}
\xi_c(r_p) &= \frac{(a^{2}+r_p^{2})\Delta'(r_p)-4r_p\Delta(r_p)} {a\,\Delta'(r_p)},\\ \eta_c(r_p) &= \frac{r_p^{2}}{a^{2}\Delta'^{\,2}(r_p)} \left[ 8\Delta(r_p)\left(2a^{2}+r_p\Delta'(r_p)\right) -r_p^{2}\Delta'^{\,2}(r_p) -16\Delta^{2}(r_p) \right],
\end{split}
\label{critical_impact_parameters}
\fe
where the prime denotes differentiation with respect to $r$. In the limit $\rho\rightarrow0$, these expressions reduce to the usual Kerr impact parameters, while the Hernquist halo modifies them through the deformation of $\Delta(r)$.

The photon shell is described by the set of spherical photon orbits whose radial coordinate lies in the interval $r_{p-}\leq r_p\leq r_{p+}$. The endpoints $r_{p-}$ and $r_{p+}$ correspond to equatorial photon rings and are obtained from
\ie\label{PO}
\eta_c(r_{p\mp})=0, \qquad \xi_c(r_{p\mp})\gtrless0.
\fe
The precise association between the signs and the prograde or retrograde branches depends on the convention adopted for $L_z$ and $\phi$. For $\rho\neq0$, the resulting algebraic equation is no longer reduced to the simple Kerr form, and the roots are more conveniently obtained numerically. Nevertheless, the Kerr limit provides a useful check,
\ie
r^{\rm Kerr}_{p\mp} = 2M\left[ 1+\cos\left( \frac{2}{3}\cos^{-1}\left(\mp\frac{a}{M}\right) \right) \right].
\fe
The intermediate spherical orbit with vanishing axial angular momentum is defined by $\xi_c(r_{p0})=0$, or equivalently
\ie
(a^{2}+r_{p0}^{2})\Delta'(r_{p0}) -4r_{p0}\Delta(r_{p0})=0.
\fe
This radius separates the prograde and retrograde sectors of the photon shell.

The polar motion is constrained by $\Theta(\theta)\geq0$. Therefore, spherical photon orbits oscillate between two turning angles $\theta_-$ and $\theta_+$, given by
\ie
\theta_{\mp}=\arccos\left(\mp\sqrt{\tilde{\omega}}\right),
\fe
where
\ie
\tilde{\omega} = \frac{ a^{2}-\eta_c-\xi_c^{2} + \sqrt{ \left(a^{2}-\eta_c-\xi_c^{2}\right)^{2} +4a^{2}\eta_c } } {2a^{2}}.
\fe
At the boundaries of the photon shell, where $\eta_c=0$, the polar oscillation shrinks to the equatorial plane. Away from these boundaries, the photons explore a finite interval in $\theta$, and the size of this interval measures how much the spherical orbit departs from a planar trajectory.

\subsection{Shadow morphology and parameter dependence}

\begin{figure}
\centering
\includegraphics[scale=0.70]{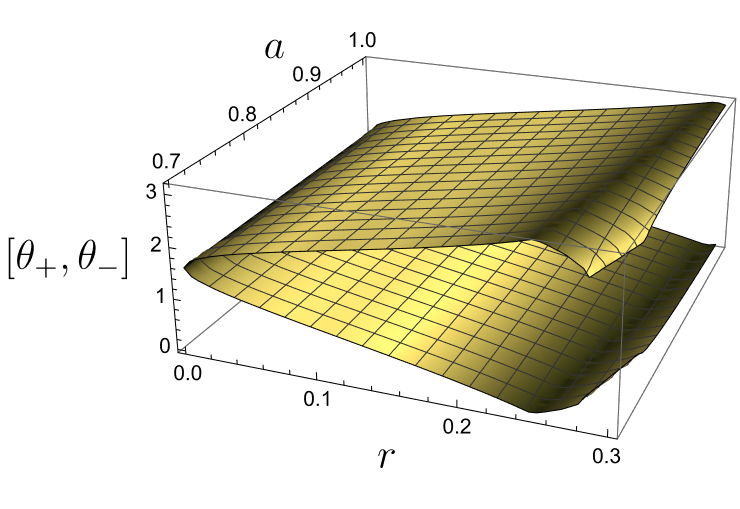}
\caption{Polar extension of the photon shell. The surfaces show the angular turning
points $\theta_+$ and $\theta_-$, together with the allowed polar interval
$\theta_+-\theta_-$, for spherical photon orbits around the rotating black
hole immersed in the Hernquist DMH. The deformation of the
surfaces reflects the combined effect of the rotation parameter $a$ and the
halo contribution encoded in $\Delta(r)$.}
\label{thetasss}
\end{figure}

We now project the critical photon orbits onto the observer's celestial plane. For a distant observer located at $r_0\rightarrow\infty$ and inclination angle $\theta_0$ with respect to the rotation axis, the celestial coordinates are
\ie
\left\{\tilde{\alpha},\tilde{\beta}\right\} = \left\{ -\xi_c\csc\theta_0,\, \pm\sqrt{ \eta_c+a^{2}\cos^{2}\theta_0 -\xi_c^{2}\cot^{2}\theta_0 } \right\}.
\label{celestial_coordinates}
\fe
The parameter $\tilde{\alpha}$ measures the apparent displacement parallel to the projected equatorial direction, while $\tilde{\beta}$ gives the vertical extension of the image. The shadow curve is obtained by varying $r_p$ over the allowed photon shell and inserting $\xi_c(r_p)$ and $\eta_c(r_p)$ into Eq.~\eqref{celestial_coordinates}.

\begin{figure}
\centering
\includegraphics[scale=0.53]{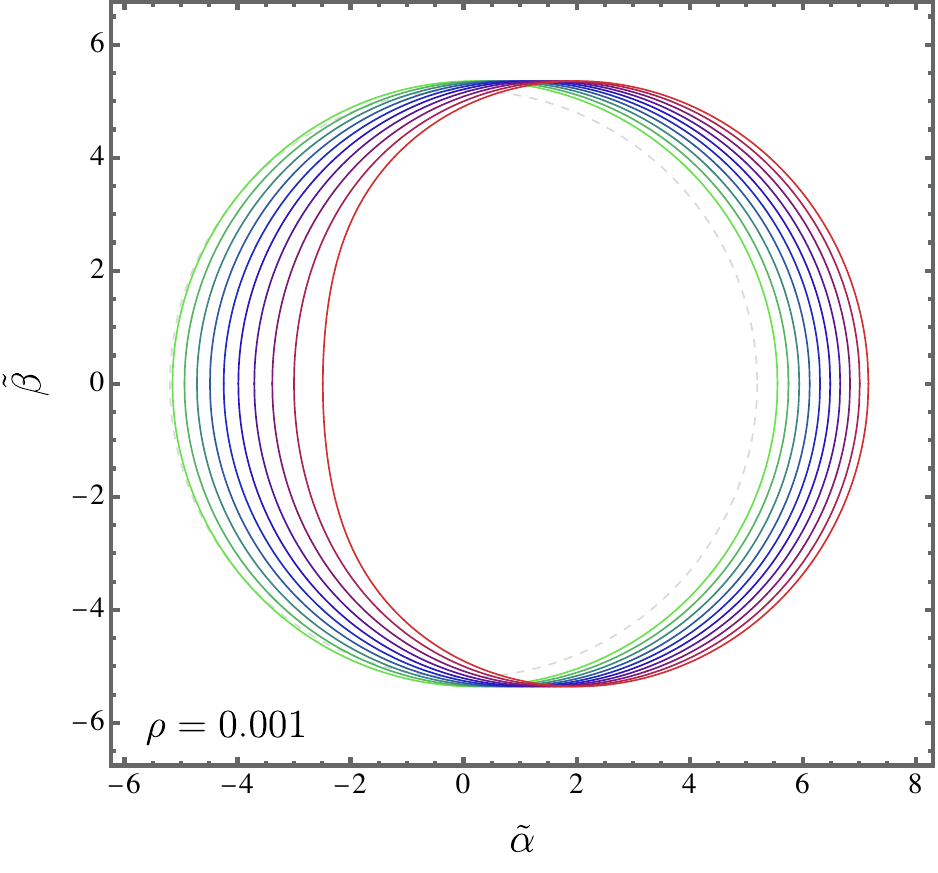}    \includegraphics[scale=0.57]{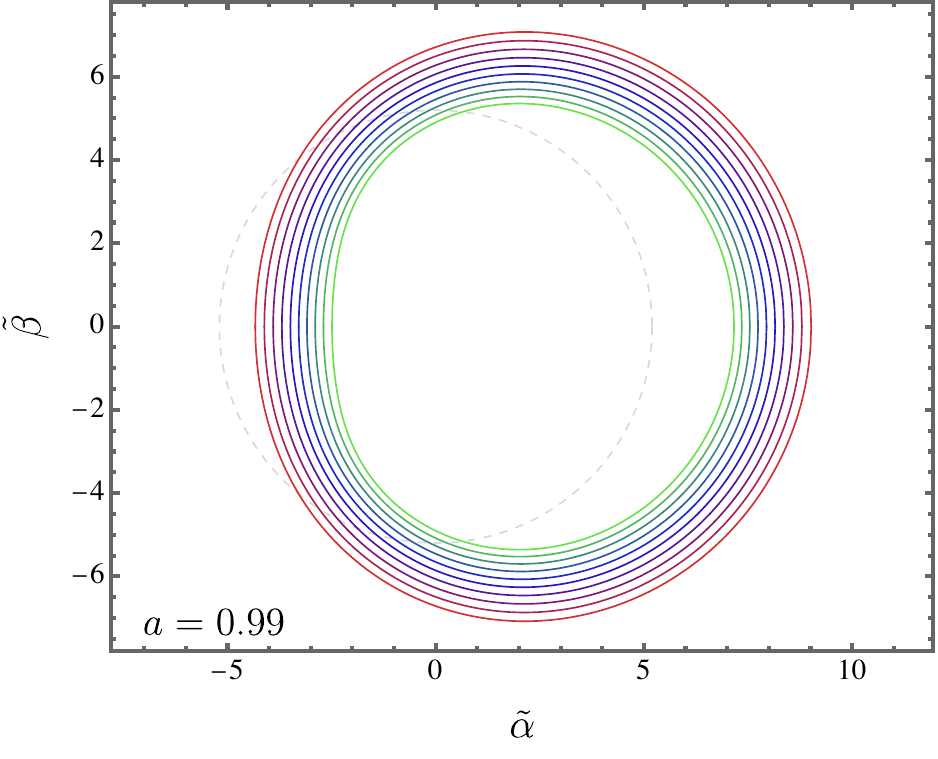}
\caption{
Shadow contours for an equatorial observer, $\theta_0=\pi/2$. The dashed
gray circle denotes the Schwarzschild reference shadow,
$R_{\rm sh}=3\sqrt{3}M$. Left panel: shadows obtained for different values
of the rotation parameter $a$ with fixed $\rho=0.001$. Increasing $a$
displaces the shadow and enhances its left--right asymmetry, producing the
characteristic rotational distortion. Right panel: shadows obtained for
different values of the Hernquist halo parameter $\rho$ with fixed
$a=0.99$.}
    \label{shadows}
\end{figure}

Figure~\ref{thetasss} displays how the allowed polar sector of the photon shell changes with the spacetime parameters. The photon shell is not merely a set of equatorial circular orbits, but a three--dimensional region of unstable null trajectories. This point is relevant because the shadow edge is produced by the full family of spherical photon orbits, and not only by the equatorial photon rings.

In Fig.~\ref{shadows}, we show the corresponding apparent shadows. For small rotation, the contour remains close to the Schwarzschild circle. As $a$ increases, frame dragging separates the prograde and retrograde photon branches, shifting the shadow horizontally and making the contour increasingly asymmetric. On the other hand, when $a$ is kept fixed and $\rho$ is increased, the Hernquist halo strengthens the effective gravitational pull felt by null rays. As a consequence, the photon capture region grows and the apparent shadow becomes larger. Notice that this behavior is consistent with the increase of the critical impact parameter produced by the halo contribution.


\subsection{EHT bounds from Sgr A$^\ast$ and M87$^\ast$}

The shadow profiles obtained above allow us to confront the rotating Hernquist black hole with horizon-scale observations. In particular, the Event Horizon Telescope (EHT) measurements of Sgr A$^\ast$ and M87$^\ast$ provide estimates for the angular diameter of the compact emission region surrounding these objects. Strictly speaking, the observed bright ring is not identical to the mathematical shadow boundary, since its precise morphology depends on the plasma distribution, the emission model, the optical depth and the radiative--transfer properties of the accretion flow.

Nevertheless, the ring diameter is strongly correlated with the photon capture region. In the absence of a source--dependent radiative--transfer treatment for the present geometry, we shall therefore use the EHT angular diameter as a phenomenological estimator of the shadow scale.

For fixed values of $(a,\rho,\theta_0)$, the shadow contour is described by the celestial coordinates given in Eq. (\ref{celestial_coordinates}). Since the contour is not necessarily circular, especially for large values of the spin parameter, it is convenient to introduce an area-equivalent shadow radius, defined by
\begin{equation}
R_{\rm sh}(a,\rho,\theta_0) = \sqrt{\frac{A_{\rm sh}}{\pi}}, \end{equation} where \begin{equation} A_{\rm sh} = 2\int_{r_p^-}^{r_p^+} \tilde{\beta}(r_p) \left| \frac{\mathrm{d}\tilde{\alpha}(r_p)}{\mathrm{d}r_p} \right| \mathrm{d}r_p .
\end{equation}
Here $r_p^-$ and $r_p^+$ denote the endpoints of the photon shell, and $R_{\rm sh}$ is measured in units of the black hole mass $M$. The corresponding theoretical angular diameter is then
\begin{equation}
\Theta_{\rm sh}^{\rm th} = 2\,R_{\rm sh}(a,\rho,\theta_0)\,\theta_g, \qquad \theta_g\equiv \frac{GM}{c^2D},
\end{equation}
where $D$ is the distance between the observer and the compact object. Throughout the numerical analysis we use the dimensionless halo parameter $\hat{\rho}=M^2\rho$. Therefore, after the rescaling $r\rightarrow r/M$ and $a\rightarrow a/M$, the radial metric function used in the bounds is
\begin{equation}
\Delta(r) = a^2+r^2-2r-\frac{32\pi \hat{\rho}\,r^2}{r+2},
\end{equation}
where we have fixed $r_s=2M$. In this parametrization, the Kerr limit is recovered for $\hat{\rho}=0$.

The observational inputs used in the comparison are summarized in Table~\ref{tab:EHTbounds}. For Sgr A$^\ast$, the EHT Collaboration reported an angular diameter $\Theta_{\rm obs}=51.8\pm2.3\,\mu{\rm as}$, whereas for M87$^\ast$ the measured diameter is $\Theta_{\rm obs}=42\pm3\,\mu{\rm as}$ \cite{EventHorizonTelescope:2019dse,EventHorizonTelescope:2022wkp,EventHorizonTelescope:2022xqj}. The angular gravitational radius $\theta_g$ is also displayed because it converts the observed angular diameter into the dimensionless diameter measured in units of $M$.

\begin{table}[t]
\centering
\begin{tabular}{c c c c}
\hline\hline
Source & $\Theta_{\rm obs}\,(\mu{\rm as})$ & $\theta_g\,(\mu{\rm as})$ 
& $\Theta_{\rm obs}/\theta_g$ \\
\hline
Sgr A$^\ast$ & $51.8\pm2.3$ & $4.8^{+1.4}_{-0.7}$ 
& $10.8^{+2.4}_{-2.8}$ \\
M87$^\ast$ & $42\pm3$ & $3.8\pm0.4$ 
& $11.1^{+2.2}_{-1.8}$ \\
\hline\hline
\end{tabular}
\caption{EHT observational inputs used to constrain the shadow size of the rotating black hole immersed in a Hernquist DMH. The last column gives the observed angular diameter in units of the angular gravitational radius.}
\label{tab:EHTbounds}
\end{table}

For each source $X=\{\mathrm{Sgr\,A}^{\ast},\mathrm{M87}^{\ast}\}$, a direct comparison with the observed angular diameter may be written as
\begin{equation}
\left| \Theta_{\rm sh}^{\rm th}(a,\hat{\rho}_X,\theta_0) - \Theta_{\rm obs}^{X} \right| \leq n\,\sigma_X,
\end{equation}
where $n=1$ and $n=2$ correspond, respectively, to the $1\sigma$ and $2\sigma$ intervals. Equivalently, one may define
\begin{equation}
\chi_X^2(a,\hat{\rho}_X,\theta_0) = \left[ \frac{ \Theta_{\rm sh}^{\rm th}(a,\hat{\rho}_X,\theta_0) - \Theta_{\rm obs}^{X} } {\sigma_X} \right]^2 ,
\end{equation}
and impose $\chi_X^2\leq n^2$.

In the present work, however, our main purpose is to obtain an upper bound on the halo contribution. This is because increasing $\hat{\rho}$ enlarges the photon capture region and, consequently, increases the area-equivalent shadow radius. Therefore, for fixed $a/M$ and $\theta_0$, values of $\hat{\rho}$ above a certain threshold overproduce the observed angular diameter. The limiting value $\hat{\rho}^{\rm max}_X$ is obtained from the upper edge of the EHT interval, $2\,R_{\rm sh}\left(a,\hat{\rho}^{\rm max}_X,\theta_0\right)\theta_g^X = \Theta_{\rm obs}^X+n\sigma_X$. In this manner, the region
$0\leq \hat{\rho}_X \leq \hat{\rho}^{\rm max}_X(a,\theta_0)$ does not exceed the angular--size constraint at the chosen confidence level. In this sense, the curves displayed below should be interpreted as upper--size bounds, rather than as a complete source--dependent likelihood analysis. A full statistical treatment would require the propagation of the uncertainties in $M$, $D$ and $\theta_g$, together with a radiative--transfer model adapted to the rotating Hernquist geometry.

\begin{figure}
\centering
\includegraphics[scale=0.59]{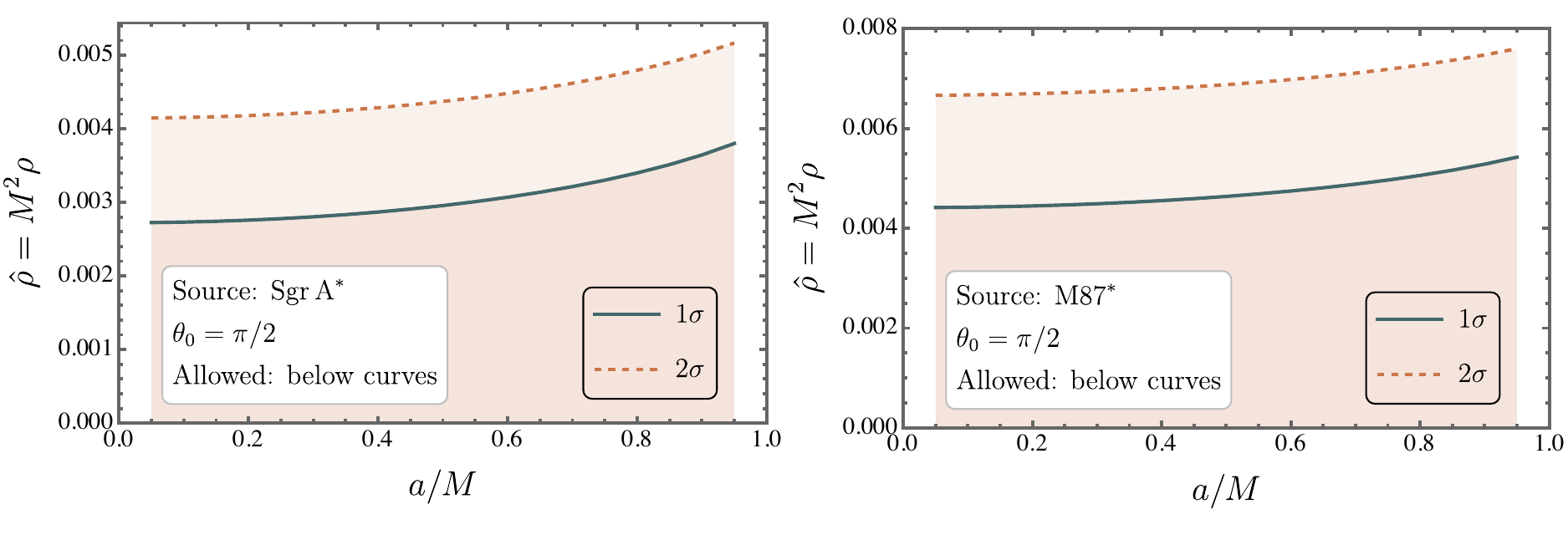}   
\caption{Upper bounds on the dimensionless Hernquist halo parameter $\hat{\rho}=M^2\rho$ obtained from the EHT angular size measurements of Sgr A$^\ast$ (left panel) and M87$^\ast$ (right panel). The solid and dashed curves correspond, respectively, to the $1\sigma$ and $2\sigma$ upper--size constraints. The shaded regions below the curves represent the parameter values that do not overproduce the observed angular diameter. The plots were generated for an equatorial observer, $\theta_0=\pi/2$, with $r_s=2M$, $0.05\leq a/M\leq0.99$ and $0\leq\hat{\rho}\leq2\times10^{-2}$.}
\label{boundsfig}
\end{figure}

For the numerical construction of Fig.~\ref{boundsfig}, we fixed an equatorial observer, $\theta_0=\frac{\pi}{2}$, and sampled the spin in the interval
\begin{equation}
0.05\leq \frac{a}{M}\leq 0.99, \qquad \Delta a=0.05 .
\end{equation}
The endpoints $a/M=0$ and $a/M=1$ were not used in the numerical grid. The first one is avoided because the critical impact parameters contain explicit factors of $1/a$, although the Schwarzschild limit can be recovered separately. The second one corresponds to the extremal boundary, where the near horizon structure requires a more careful numerical treatment. For the halo parameter we used
\begin{equation}
0\leq \hat{\rho}\leq 2\times 10^{-2}, \qquad \Delta \hat{\rho}=2.5\times 10^{-4},
\end{equation}
and the shadow area was evaluated from a polygonal approximation of the contour generated with $350$ points along the photon-shell interval. The values of $\hat{\rho}^{\rm max}_X$ were then obtained by linear interpolation between neighboring points of the numerical grid.

Using the central values of $\theta_g$ listed in Table~\ref{tab:EHTbounds}, the upper target radii entering the numerical bounds are
\begin{equation}
R_{\rm up}^{\mathrm{Sgr\,A}^{\ast}} = \frac{51.8+n(2.3)}{2(4.8)} =
\begin{cases}
5.64, & n=1,\\
5.88, & n=2,
\end{cases}
\end{equation}
and
\begin{equation}
R_{\rm up}^{\mathrm{M87}^{\ast}} = \frac{42+n(3)}{2(3.8)} =
\begin{cases}
5.92, & n=1,\\
6.32, & n=2.
\end{cases}
\end{equation}
These numbers explain why M87$^\ast$ gives a less restrictive upper bound on $\hat{\rho}$ in Fig.~\ref{boundsfig}: its allowed dimensionless upper radius is larger than the corresponding value for Sgr A$^\ast$. In addition, the curves slowly increase with $a/M$. This happens because, for an edge on observer, the Kerr--like spin deformation tends to reduce the area equivalent radius of the shadow when compared with the nonrotating reference. As a result, a larger halo contribution can be tolerated before the theoretical angular diameter reaches the EHT upper limit.

From Fig.~\ref{boundsfig}, one sees that Sgr A$^\ast$ provides the tighter constraint. For the range of spins explored here, the $1\sigma$ curve gives approximately $\hat{\rho}^{\rm max}_{\mathrm{Sgr\,A}^{\ast}}\sim(2.7-3.8)\times10^{-3}$, while the $2\sigma$ curve gives $\hat{\rho}^{\rm max}_{\mathrm{Sgr\,A}^{\ast}}\sim(4.1-5.2)\times10^{-3}$. For M87$^\ast$, the corresponding bounds are weaker, with $\hat{\rho}^{\rm max}_{\mathrm{M87}^{\ast}}\sim(4.4-5.4)\times10^{-3}$ at $1\sigma$ and $\hat{\rho}^{\rm max}_{\mathrm{M87}^{\ast}}\sim(6.7-7.6)\times10^{-3}$ at $2\sigma$. Therefore, both sources restrict the amount of Hernquist dark matter that can be placed around the compact object without producing a shadow larger than the observed EHT ring.

Finally, the bounds obtained from Sgr A$^\ast$ and M87$^\ast$ should not be combined into a single universal constraint on $\rho$. The parameter $\rho$ describes the local density scale of the Hernquist distribution and may depend on the astrophysical environment of each compact object. Nevertheless, the two sources test the same geometrical mechanism: the halo term enlarges the photon capture region, while the EHT angular size measurements restrict the magnitude of this enlargement.


\section{Lensing phenomena}
Gravitational lensing refers to the bending of light rays as they move through a gravitational field. This effect can be understood in two distinct regimes. In the weak-field regime, light passes at large distances from the massive object (the lens), where gravity only slightly alters its path. In contrast, the strong-field regime occurs when light travels very close to the compact object, where the gravitational field is extremely intense, and the deflection angle can grow without bound as the light approaches a critical distance.

\subsection{Deflection Angle in Strong Field Regime and Observables} In this subsection, we focus on how light is deflected by gravity in the strong gravitational regime when its motion is confined to the equatorial plane ($\theta=\pi/2$). For simplicity, we assume that both the light source and the observer are located far away from the black hole, in regions where the gravitational field becomes negligible \cite{Bozza:2002zj,Virbhadra:1999nm,Perlick:2003vg}. This assumption not only simplifies the analysis but also makes it possible to derive analytical expressions. Restricting the motion to the equatorial plane further reduces the mathematical complexity, while still capturing the essential physics and yielding results that can be extended to more general configurations.  

\begin{figure}
    \centering
    \begin{tabular}{ccc}
     \includegraphics[width=0.5\textwidth]{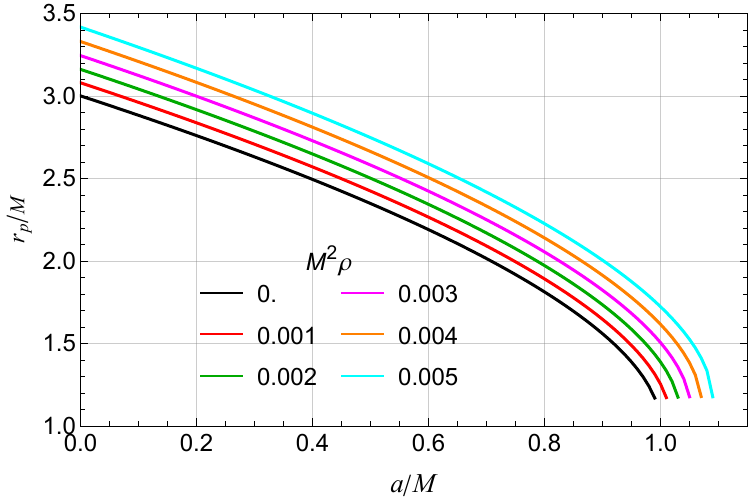}&
     \includegraphics[width=0.5\textwidth]{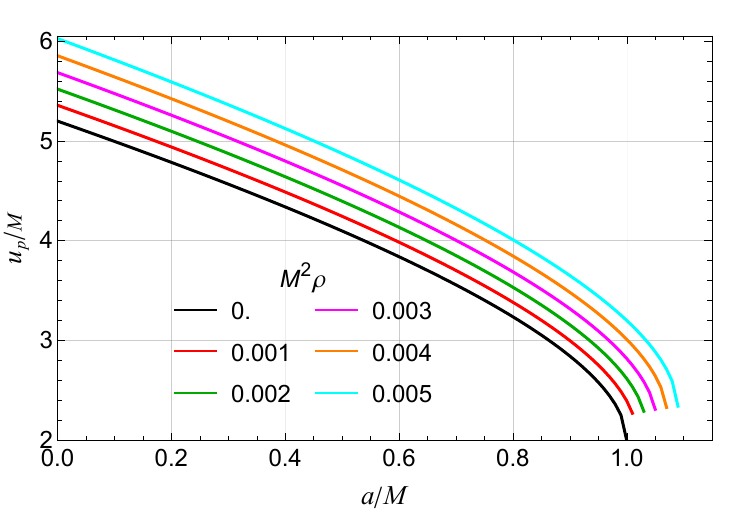}
     \end{tabular}
    \caption{\textbf{Left:} Unstable circular photon orbit radius $r_p/M$ versus $a/M$ for selected DMH parameter values $\hat{\rho}$. \textbf{Right:} Corresponding critical impact parameter $u_p/M$ as a function of $a/M$}
    \label{photonorbits}
\end{figure}

Let us consider a photon that starts from an asymptotically flat region ($r\to\infty$) and moves towards the black hole until it reaches its minimum radial distance from the centre of the black hole, $r_0$. This location marks the turning point of its trajectory, where the intense gravitational field alters its direction, allowing the photon to escape and continue in another asymptotically flat region. This turning point satisfies the mathematical condition $\dot{r}|_{r=r_0}=0$ or $\mathcal{V}_{\pm}(r=r_0)=E$, which leads us to the following expression of impact parameter
\begin{equation} \label{angmom}
\frac{L}{E}=u(r_0) = \frac{\pm r_0 (r_0+2) \sqrt{a^2+\frac{r_0 \left(r_0^2-32 \pi  \hat{\rho}  r_0-4\right)}{r_0+2}}-2 a (16 \pi  \hat{\rho}  r_0+r_0+2)}{r_0^2-32 \pi  \hat{\rho}  r_0-4}.
\end{equation}Photon trajectories depend on their direction of motion relative to the black hole spin. Prograde and retrograde photons follow distinct paths; here, we restrict to the positive branch of Eq.~(\ref{angmom}), corresponding to counterclockwise motion. As the impact parameter decreases, the deflection angle increases and diverges at a critical value, where photons execute circular motion, forming the \textit{photon sphere}. These orbits are unstable under radial perturbations. The dependence of the photon sphere radius $r_{p}$ for prograde photons (that can be obtained from Eq. \eqref{PO}) on spin $a$ for different values of $\hat{\rho}$ is shown in Fig.~\ref{photonorbits} (left panel), where $r_{p}$ decreases with increasing $a$, whereas it increases with $\hat{\rho}$. The minimum impact parameter $u_p=u|_{r=r_p}$ (right panel of Fig.~\ref{photonorbits}) also exhibits similar trends to $r_{p}$ with $a$ and $\hat{\rho}$.

We now examine the deflection angle and the associated lensing coefficients in the strong-field regime ($r_0 \to r_{p}$) for photons in the considered spacetime, with particular emphasis on the influence of the spacetime parameters. We can write the angle of deflection for the considered spacetime, as a function of the distance of closest approach $r_0$, in the following manner \cite{Bozza:2002zj}
\begin{eqnarray}\label{bending1}
\alpha_{D}(r_0)=I(r_0)-\pi,
\end{eqnarray} 
with

\begin{eqnarray}\label{bending2}
I(r_0) = 2 \int_{r_0}^{\infty}\frac{\mathrm{d}\phi}{\mathrm{d}r} \mathrm{d}r = 2\int_{r_0}^{\infty}\frac{\sqrt{A(r_0) B }\left(2AL+ D\right)}{
\sqrt{4AC+D^2}\sqrt{A(r_0) C-A C(r_0)+L\left(AD(r_0)-A(r_0)D\right)}} \mathrm{d}r.
\end{eqnarray} 
It is well known that the deflection angle becomes divergent as $r_0 \to r_{m}$, which makes a direct evaluation of the integral impractical. To overcome this issue, one typically expands the integrand in a series around the photon sphere radius \cite{Virbhadra:1999nm, Claudel:2000yi, Bozza:2002zj}. Following the method developed by Tsukamoto \cite{Tsukamoto:2016jzh}, which builds on and improves Bozza’s earlier approach \cite{Bozza:2002zj}, we introduce a new variable $z = 1 - r_0/r_{p}$ to isolate the dominant contribution near the divergence.
With this treatment, one obtains an analytic expression for the deflection angle in the strong-field regime, written in terms of the impact parameter $u$ as:
\begin{eqnarray}\label{sdef2}
\alpha_{D}(u) &=& \bar{a} \log\left(\frac{u}{u_{p}} - 1\right) + \bar{b} + \mathcal{O}(u - u_{p}),
\end{eqnarray}
\begin{figure*}
	\begin{centering}
		\begin{tabular}{ccc}
		    \includegraphics[width=0.5 \textwidth]{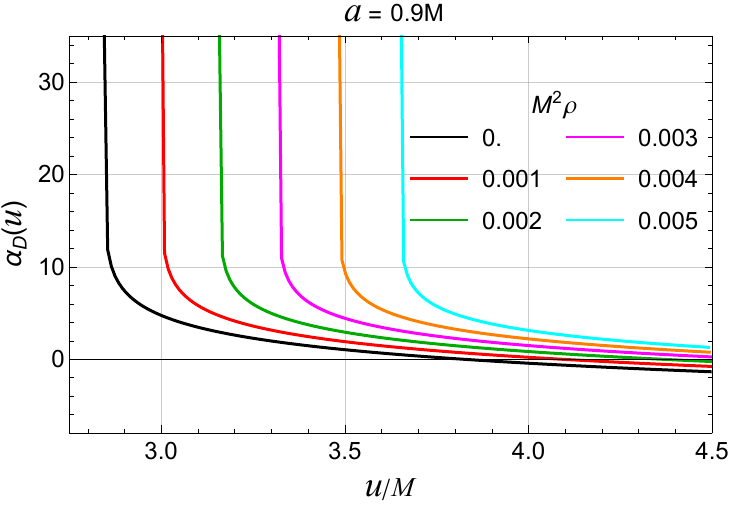}
		    
		    \end{tabular}
	\end{centering}
	\caption{Deflection angle $\alpha_D$ versus impact parameter $u$ for the new axisymmetric black holes immersed in  Hernquist DMH (colored curves) compared to Kerr ($\hat{\rho}=0$, black). The logarithmic divergence near $u_p$ (inset) depends on  $\hat{\rho}$.}\label{deflection}		
\end{figure*} where $\bar{a}$ and $\bar{b}$ are the strong-field lensing coefficients. We used numerical methods to get $\bar{a}$ and $\bar{b}$ and hence the deflection angle, which can be seen in Fig. \ref{deflection}. It is clear from the Fig. \ref{deflection} that the Hernquist DMH parameter $\hat{\rho}$ increases the deflection of light around the black holes.
Next, we turn to the observable features of strong gravitational lensing, which are derived from the lens equation in combination with the photon deflection angle. The lens equation provides the geometric relation between the observer $O$, the lens $L$, and the source $S$. Assuming that both the observer and the source are located far from the lens, are nearly aligned with it, and that the surrounding spacetime is effectively flat \cite{Bozza:2002zj, Bozza:2008ev, Virbhadra:1999nm}, the lens equation can be written as  
\begin{eqnarray}\label{lenseq}
\beta &=& \theta - \mathcal{D}\,\Delta \alpha_n,
\end{eqnarray}where $\Delta \alpha_n = \alpha(\theta) - 2n\pi$ represents the small residual deflection after the photon completes $2n\pi$ loops around the lens. Here, $n$ is a positive integer, and the analysis is restricted to the regime $0 < \Delta \alpha_n \ll 1$. The angles $\beta$ and $\theta$ correspond to the angular positions of the source and the image relative to the optical axis. The distances between the observer, lens, and source are denoted by $D_{\text{OL}}$, $D_{\text{LS}}$, and $D_{\text{OS}} = D_{\text{OL}} + D_{\text{LS}}$, respectively, with $\mathcal{D} = D_{\text{LS}}/D_{\text{OS}}$.

To determine the angular position of the $n$-th relativistic image, we substitute Eq.~(\ref{sdef2}) into the lens equation (\ref{lenseq}), which leads to \cite{Bozza:2002zj}  
\begin{equation}
\theta_n = \theta_n^0 + \frac{u_{p}\, e_n\, (\beta - \theta_n^0)\, D_{\text{OS}}}{\bar{a}\, D_{\text{LS}}\, D_{\text{OL}}},
\end{equation}

where  
\begin{equation}
e_n = \exp\!\left(\frac{\bar{b}}{\bar{a}} - \frac{2n\pi}{\bar{a}}\right), 
\qquad 
\theta_n^0 = \frac{u_p (1 + e_n)}{D_{\text{OL}}}.
\end{equation}Here, $\theta_n^0$ denotes the image position corresponding to an exact deflection of $2n\pi$. In the special case of perfect alignment ($\beta = 0$), the images appear as concentric Einstein rings with angular radii  
\begin{equation}\label{einstein-rings}
\theta_n^E = \frac{u_p}{D_{\text{OL}}} (1 + e_n).
\end{equation}The gravitational bending of light alters the cross-sectional area of the light beam, resulting in a magnification of the observed image. According to Liouville’s theorem, the surface brightness remains unchanged during lensing, so the magnification depends solely on the ratio of the solid angles of the image and the unlensed source. Consequently, the magnification of the $n$-th relativistic image is given by \cite{Virbhadra:1999nm, Bozza:2002zj}  
\begin{equation}\label{mag}
\mu_n = \left(\frac{\beta}{\theta}\frac{d\beta}{d\theta}\right)^{-1}\Bigg|_{\theta_n^0}
= \frac{u_p\, e_n (1 + e_n)\, D_{\text{OS}}}{\bar{a}\, \beta\, D_{\text{LS}}\, D_{\text{OL}}^2}.
\end{equation}For the limiting case where the source is perfectly aligned with the optical axis ($\beta \to 0$), Eq.~(\ref{mag}) becomes divergent. This reflects the fact that gravitational lensing is most prominent under exact alignment conditions. In addition, the magnification decreases exponentially with increasing image order $n$, which means that the first relativistic image $\theta_1$ is the brightest among them. Since the outermost image ($n=1$) is well separated from the closely packed sequence of higher-order images ($n=2,3,\ldots,\infty$), it is convenient to introduce the following observable quantities \cite{Bozza:2002zj}:
\begin{eqnarray}\label{observable}
    && \theta_{\infty}=\frac{u_p}{D_{OL}},
    ~~~~~ s=\theta_1-\theta_{\infty}\approx \theta_{\infty} \exp\left[\frac{\bar{b}-2\pi}{\bar{a}}\right],
    \nonumber\\
    && r_\text{mag}=\frac{\mu_1}{\sum^{\infty}_{n=2} \mu_{n}}\approx\frac{5\pi}{\bar{a}\log(10)}. 
\end{eqnarray}Here, $\theta_{\infty}$ denotes the limiting angular position at which the infinite sequence of higher-order images accumulates. The quantity $s$ measures the angular separation between the first relativistic image and the remaining cluster of images. Meanwhile, $r_\text{mag}$ represents the ratio of the flux from the first image to the combined flux of all the other, more tightly grouped images.
\begin{figure*}
	\begin{centering}
		\begin{tabular}{ccc}
		    \includegraphics[width=0.5 \textwidth]{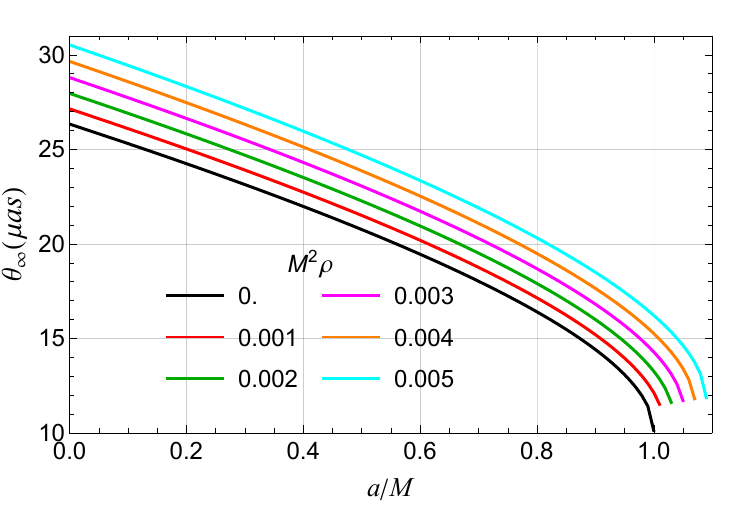}&
		    \includegraphics[width=0.5 \textwidth]{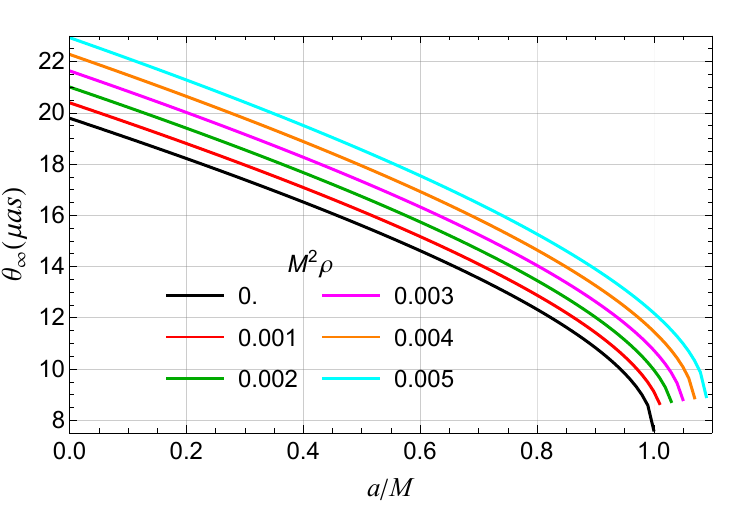}\\
            \includegraphics[width=0.5 \textwidth]{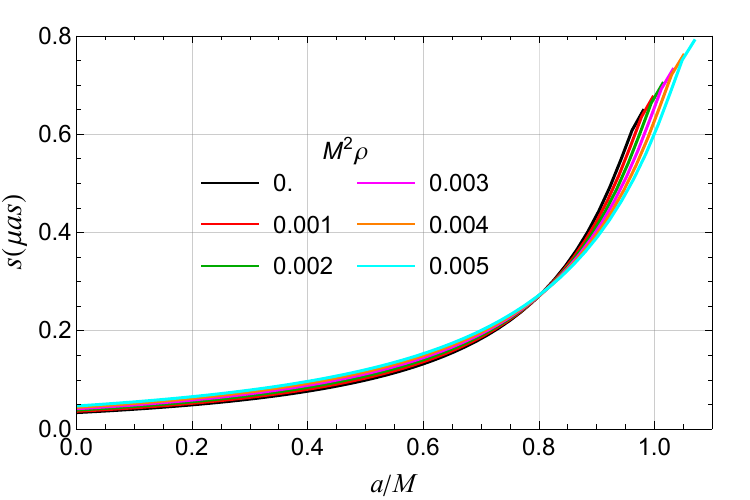}&
		    \includegraphics[width=0.5 \textwidth]{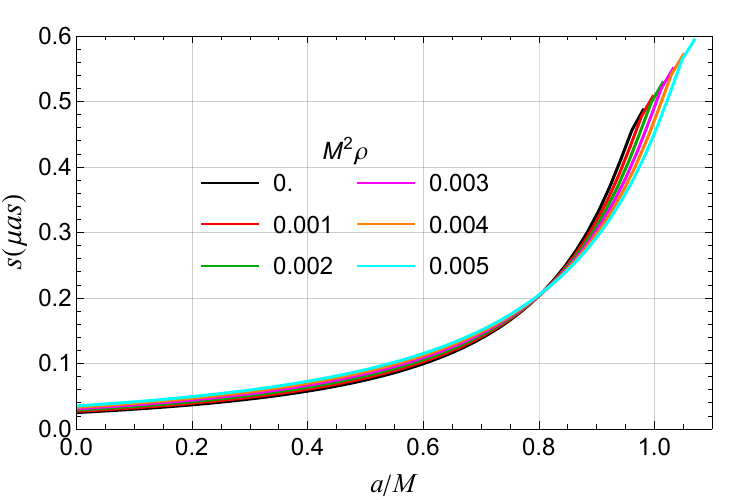}
		    \end{tabular}
	\end{centering}
	\caption{Strong lensing observables for supermassive black holes Sgr\,A* (left) and M87* (right). \textbf{Top:} Angular position $\theta_\infty$ of the photon ring versus spin $a$ for different DMH parameter $\hat{\rho}$. \textbf{Bottom:} Angular separation $s$ between first and higher-order images versus $a$ for different DMH parameter $\hat{\rho}$. }\label{observables}		
\end{figure*} 
\begin{figure*}
	\begin{centering}
		\begin{tabular}{ccc}
		    \includegraphics[width=0.5 \textwidth]{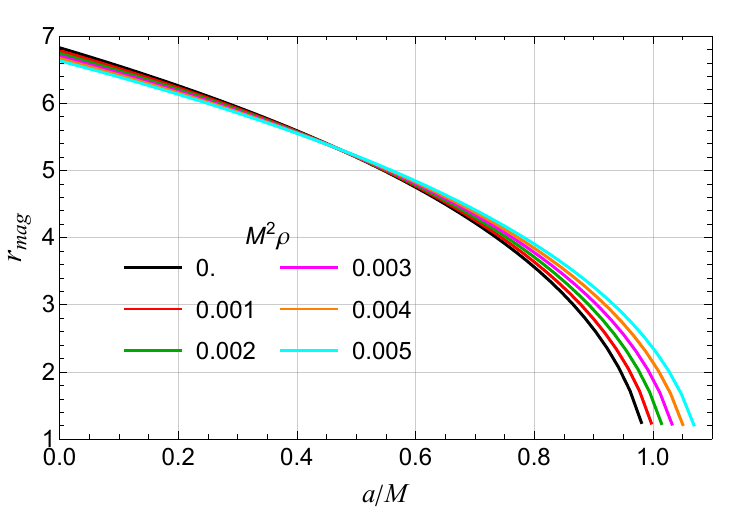}&
    		    \end{tabular}
	\end{centering}
	\caption{Flux magnification ratio $r_{\text{mag}}$ as a function of spin $a$ for different DMH parameter $\hat{\rho}$.}\label{plotrmag}		
\end{figure*}

Another important observable in the strong-deflection regime is the time delay between relativistic images. Such delays arise when light rays undergo large deflections, with the bending angle exceeding $2\pi$, allowing photons to circle the black hole one or more times before reaching the observer. Since different relativistic images are formed by photons following distinct trajectories around the lens, the corresponding optical path lengths and travel times are generally different. As a result, a measurable time delay is produced between the arrival of these images. For two relativistic images, labeled by $p$ and $q$, that appear on the same side of the lens, the corresponding time delay is given by \cite{Bozza:2003cp}
\begin{eqnarray}\label{TD2}
\Delta T_{p,q} \approx 2\pi(p-q) u_p. 
\end{eqnarray}
Therefore, the time delay between any two relativistic images formed on the same side of the lens is directly governed by the critical photon impact parameter, $u_p$. Precise measurements of this delay can provide valuable information about the lensing system, offering an advantage over methods that rely solely on the system's angular or physical dimensions. In particular, time-delay observations can be used to determine the distance to the black hole with high accuracy. Such measurements require the background source to exhibit intrinsic variability; however, this condition is not especially restrictive, as variable stars and other time-varying astrophysical sources are commonly found in galaxies.

\begin{table*}[!ht]
 \resizebox{\textwidth}{!}{ 
	\begin{tabular}{|c| c| c c c| c c c| c| }
\hline
\multicolumn{1}{|c|}{}&
\multicolumn{1}{c|}{}&
\multicolumn{3}{c|}{Sgr A*}&
\multicolumn{3}{c|}{M87*}& \\
{$a$ }&\;\ {$\hat{\rho}$}  & {$\theta_\infty $ ($\mu$as)} & {$s$ (nas) } & $\Delta T_{2,1}$(min) & {$\theta_\infty $ ($\mu$as)} & {$s$ (nas) } &$\Delta T_{2,1}$(hrs)&\;\; {$r_{mag}$ } 
\\ \hline
\multirow{6}{*}{0} &0  & 26.3299 & 0.0329517 & 11.49676 & 19.7820 & 0.02475713 & 289.647 & 6.82188 \\ 
& 0.001 & 27.1329& 0.0354741& 11.8474 & 20.3853& 0.0266522& 298.481 &  6.77905\\ 
  & 0.002 & 27.9534 & 0.0380856& 12.2057 & 21.0018& 0.0286143& 307.508 & 6.73833 \\
 & 0.003 & 28.7913& 0.0407799& 12.5716 & 21.6314& 0.0306385& 316.725   & 6.6997 \\ 
  & 0.004 & 29.6462& 0.0435499&  12.9448 & 22.2736& 0.0327197& 326.13 &  6.66313 \\ 
   & 0.005 & 30.5177& 0.0463884& 13.3254 & 22.9284& 0.0348523& 335.716 & 6.62858 \\ 
\hline 
\multirow{6}{*}{0.3} &0 & 23.1327& 0.0604049& 10.1008 & 17.38& 0.0453831& 254.477 & 5.93812 \\
& 0.001 & 23.8998& 0.0639604& 10.4357 & 17.9562& 0.0480544& 262.914 & 5.91804 \\ 
  & 0.002 & 24.6863& 0.0675829& 10.7791 & 18.5472& 0.0507761& 271.567 & 5.89938 \\
 & 0.003 & 25.4921& 0.0712618& 11.131 & 19.1526& 0.0535401& 280.431 &  5.88214 \\ 
  & 0.004 & 26.3168& 0.0749861& 11.491 & 19.7722& 0.0563382& 289.503 &  5.8663 \\ 
   & 0.005 & 27.16& 0.0787449& 11.8592 & 20.4057& 0.0591622& 298.779 &  5.85184 \\  
\hline 
\multirow{6}{*}{0.6} &0 & 19.4504& 0.131981& 8.49288 & 14.6133& 0.0991593& 213.968 &  4.74756 \\ 
& 0.001 & 20.1864& 0.136166& 8.81428 & 15.1664& 0.102304& 222.065 & 4.76626 \\ 
  & 0.002 & 20.9443& 0.140353& 9.1452 & 15.7358& 0.105449& 230.402 & 4.78485 \\
 & 0.003 & 21.7237& 0.144526& 9.48552 & 16.3214& 0.108585& 238.976 &  5.39317 \\ 
  & 0.004 & 22.5243& 0.148673& 9.83508 & 16.9228& 0.1117& 247.783 &  4.8219 \\ 
   & 0.005 & 23.3456& 0.152781& 10.1937 & 17.5399& 0.114786& 256.818 &  4.84043 \\ 

\hline
\end{tabular}}
\caption{The lensing observables $\theta_\infty $ ($\mu$as), $s$ ($\mu$as), time delay $\Delta T_{2,1}$, and $r_{mag}$ for massive black holes Sgr A* and M87.
\label{table1} }
\end{table*} 

Further, we want to model the supermassive black holes M87* and Sgr A* as the new axis-symmetric black holes immersed in Hernquist DMH and studied their lensing observables. We then compare these results with those obtained for the standard Kerr black holes. Current astronomical observations estimate the mass of M87* to be $\left(6.5\pm0.7\right)\times{10}^9M_\odot$, located at a distance of $16.8$ MPc \cite{EventHorizonTelescope:2019dse}. For Sgr A*, the estimated mass is $4.28_{\pm0.10}^{\pm0.21}\times{10}^6M_\odot$ with a distance of $8.32^{\pm0.07}_{\pm0.14}$ KPc \cite{2017ApJ...837...30G}. The relativistic angular image position, $\theta_{\infty}$, decreases monotonically with spin $a$, whereas it increases with DMH parameter $\hat{\rho}$. The behaviour of the angular separation observable $s$ shows a clearer difference between the models. We find that the new axis-symmetric black holes immersed in Hernquist DMH produce larger values of $s$ compared to Kerr black holes for slowly rotating black holes, whereas this trend reverses for higher values of $a$ (see Fig.~\ref{observables}). The relative flux ratio $r_{mag}$ behaves in the opposite way to $s$, which can be confirmed through Fig.~\ref{plotrmag}. Numerical values of strong lensing observables, $\theta_\infty$, $s$, and $r_{mag}$ for the new axis-symmetric black holes immersed in Hernquist DMH are listed in Table \ref{table1}.

 \subsection{Deflection Angle in Weak Field Regime and Einstein Ring Size}
 Gravitational lensing in the weak-field domain is among the most well-established and experimentally verified consequences of GR \cite{Weinberg:1972kfs, Bartelmann:2010fz}. When the gravitational field is sufficiently weak, the bending of light can be determined through a perturbative treatment of the spacetime metric, leading to the standard post-Newtonian formulas that accurately account for lensing phenomena produced by astrophysical objects such as stars, galaxies, and galaxy clusters. Within this framework, the new axis-symmetric black holes immersed in Hernquist DMH are expected to introduce subtle corrections to the conventional light-deflection predictions of GR \cite{Berti:2015itd, Cardoso:2019rvt}. Exploring these effects is particularly important in view of the rapidly increasing precision of modern astrometric observations. Instruments such as VLTI/GRAVITY, together with upcoming observational facilities, offer the possibility of detecting small deviations from classical GR expectations through lensing measurements. Motivated by this prospect, we derive the weak-field deflection angle associated with the new axis-symmetric black holes immersed in Hernquist DMH spacetime described by metric (\ref{NSR}) and examine its impact on the properties of the Einstein ring.

\begin{figure*}
	\begin{centering}
		\begin{tabular}{ccc}
		    \includegraphics[width=0.5 \textwidth]{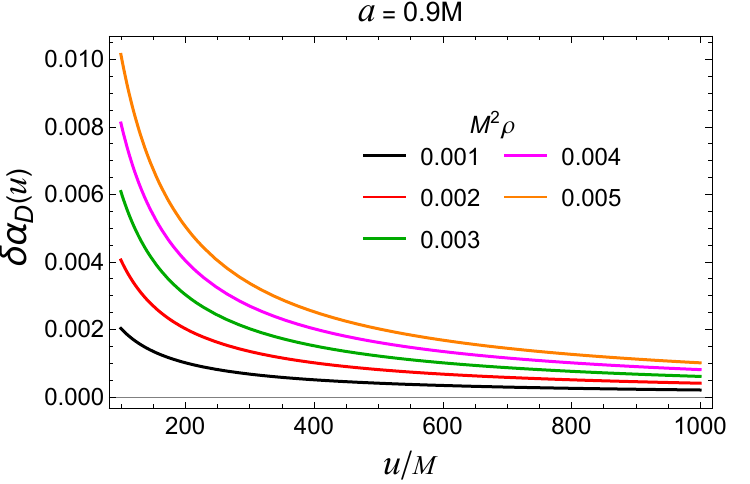}
    		    \end{tabular}
	\end{centering}
	\caption{The deviation in weak lensing deflection angle $\delta\alpha_{D}(u)=\alpha_D(u)-\alpha_D(u)|_{Kerr}$ vs impact parameter $u$ of the new axisymmetric black holes immersed in Hernquist DMH.}\label{plotwkdef}
    \end{figure*}

 To derive the light-bending angle, we consider a photon that travels from an asymptotically flat region toward the black hole. Along its trajectory, the photon reaches a minimum radial coordinate $r_0$, referred to as the turning point, with $r_0$ lying outside the photon sphere ($r_0	
 >r_{\text{ps}}$). The photon is subsequently deflected by the gravitational field and escapes to a distant asymptotically flat region on the opposite side of the lens without making full or multiple loops around the black hole. Instead, they experience only a small bending due to the black hole’s gravity, leading to a deflection angle less than $2\pi$. Next, we take taylor expansion of the integrand of the deflection angle expression \eqref{bending2} by defining a new variable, $z=r_0/r$, and then we integrate to get the deflection angle in the weak field regime of the new axis-symmetric black holes immersed in Hernquist DMH as
 
\begin{eqnarray}\label{bending3}
       \alpha(r_0)&=&\frac{4(1+16 \pi  \hat{\rho}) }{r_0}+\frac{1}{r_0^2}\left[-4(1+a)+\frac{15 \pi }{4}+8 \pi \hat{\rho} \left(9 \pi -8 (a+2)+8 \pi \hat{\rho}(15 \pi -16)\right)\right]\nonumber\\&&+\frac{1}{r_0^3}\Bigg[\left(-\frac{15\pi}{2}+\frac{122}{3}\right)+2a\left(8-5\pi+a\right)+8\pi\hat{\rho}\Bigg(514-33\pi+4a(128-8\pi+a)\nonumber\\&&-8\pi\hat{\rho}\left(66(\pi-4)+8a(5\pi-8)+32\pi\hat{\rho}\left(\frac{244}{3}-15\pi\right)\right)\Bigg)\Bigg].
   \end{eqnarray}
   It is evident from the above result that the deflection angle decreases as the closest approach distance $r_0$ moves farther from the black hole. To rewrite the deflection angle as a function of the impact parameter $u$, we begin by expanding $1/r_0$ as a power series in $1/u$, which yields
   \begin{eqnarray}\label{expansion}
     \frac{1}{r_0}=\frac{1}{u}+\frac{1-16 \pi  \hat{\rho}}{u^2}+\frac{a^2-4 a (1+16 \pi  \hat{\rho} )+5 (1-16 \pi  \hat{\rho} )^2}{2u^3}+\mathcal{O}\left(\frac{1}{u^4}\right).  
   \end{eqnarray}Upon inserting Eq. (\ref{expansion}) into Eq. (\ref{bending3}), the weak deflection angle takes the following form in terms of the impact parameter $u$
  \begin{eqnarray}\label{bending4}
       \alpha(b)&=&\frac{4(1+16\pi\hat{\rho})M}{b}+\frac{M}{4b^2}\Bigg[15\pi M -16a+8\pi\hat{\rho}\Bigg(\left(-16+9\pi\right)M-8a+8\pi\hat{\rho} M\left(
       60\pi-92\right)\Bigg)\Bigg]\nonumber\\&&+\frac{M}{3b^3}\Bigg[128M^2+6a\left(2a-5\pi M\right)+16\pi\hat{\rho}\Bigg(\left(208-45\pi\right)M^2+12a\left(4M(\pi-1)+a\right)\nonumber\\&&+16\pi\hat{\rho}\Big(\left(192-54\pi\right)M^2-6aM\left(5\pi-8\right)+16\pi\hat{\rho} M^2\left(196-45\pi\right)\Big)\Bigg)\Bigg]+\mathcal{O}\left(\frac{1}{b^4}\right),
   \end{eqnarray}where $b=uM$ represents the rescaled impact parameter. In the absence of DMH, one can easily get the deflection angle of Kerr black holes, which reads \cite{Ovgun:2019wej,Li:2021lhr,He:2022yco,Kumar:2025bim}
   \begin{eqnarray}\label{bendingkerr}
       \alpha(b)&=&\frac{4M}{b}+\frac{15\pi M^2}{4b^2} -\frac{4aM}{b^2}+\frac{128M^3}{3b^3}+\frac{4a^2M}{b^3}-\frac{10\pi aM^2}{b^3}+\mathcal{O}\left(\frac{1}{b^4}\right),
   \end{eqnarray}which further reduces to weak lensing deflection of Schwarzschild black holes \cite{KumarWalia:2022ddq, Kumar:2025bim,Kumar:2025ueq}, when we switch off the spin ($a=0$). Fig.~\ref{plotwkdef} illustrates the deviation of the weak-field deflection angle for the new axisymmetric black holes immersed in a Hernquist DMH from the corresponding Kerr black hole prediction. The positive nature of this deviation indicates that the presence of the DMH enhances the bending of light. Furthermore, the Fig.~\ref{plotwkdef} clearly shows that this enhancement becomes more pronounced as the DMH density parameter $\hat{\rho}$ increases, highlighting the growing influence of the surrounding dark matter distribution on the lensing behavior.

   Now, we want to inspect the effect of DMH density $\hat{\rho}$ on the size of the Einstein Rings of the rotating black holes. In this study, we focus on the idealized lensing configuration in which the source, lens, and observer are perfectly aligned, while both the source and the observer are located in regions where spacetime is effectively flat. To investigate the Einstein ring, we employ observational data from the galaxy ESO325-G004, whose total mass is estimated to be $M = 1.50 \times 10^{11} M_{\odot}$ \cite{Smith:2005ru, Smith:2013ena}, with $M_{\odot} = 1.98 \times 10^{30}\, {\rm kg}$ denoting the solar mass. This mass accounts for contributions from the central black hole as well as the galaxy’s luminous and dark matter components. The lensed background galaxy is observed at a redshift of $z_s=1.141$. To determine the relevant lensing distances, we make use of Hubble’s law, $$cz = H_{0} D,$$ where $H_0=1.542\times10^2\,{\rm Mpc}$ is the Hubble constant and $D$ represents the proper distance. Assuming a spatially flat universe, the comoving distance is related to the proper distance through $$d=D(1+z).$$ Applying these relations to the observed redshifts allows us to calculate the distances among the observer, lens, and source that are required for the lensing analysis as
   \begin{align}
D_{\text{LS}}=\dfrac{c z_s(1+z_s)}{H_0}=2.863\times10^4\,{\rm Mpc}, \nonumber\\ D_\text{OL}=\dfrac{c z_l(1+z_\ell)}{H_0}=1.542\times10^2\,{\rm Mpc} \ .\label{ESOdata}
\end{align} For the lens galaxy ESO325-G004 at redshift $z_l=0.0345$, the angular size of the Einstein ring has been reported as \cite{Smith:2005ru,Smith:2013ena}
\begin{equation}\label{obsering}
\theta_{\text{E}}^{\rm obs} = (2.85^{+0.55}_{-0.25})'' .
\end{equation}
In a comprehensive study of gravitational lens equations, Bozza \textit{et al.} \cite{Bozza:2008ev} showed that the Ohanian lens equation and its related variants offer highly accurate approximations to the exact lensing geometry. They further recast the Ohanian equation into a form involving the distances between the observer, lens, and source planes, yielding the expression given below
   \begin{align}
D_{OS}\tan\beta=\dfrac{D_{OL}\sin\theta-D_{LS}\sin(\alpha-\theta)}{\cos(\alpha-\theta)}.\label{LE}
\end{align}Here, $\beta$ and $\theta$ denote the angular positions of the unlensed source and the lensed image, respectively. The distances from the observer to the lens plane and from the lens plane to the source plane are represented by $D_\text{OL}$ and $D_\text{LS}$, with $D_\text{OS}=D_\text{LS}+D_\text{OL}$. The impact parameter is related to the image angle through $\sin\theta=b/D_\text{OL}$.

When the source, lens, and observer lie on the same line of sight, one has $\beta=0$, and the image appears as an Einstein ring of angular radius $\theta_E$. In this perfectly aligned configuration, both $\alpha_\text{D}$ and $\theta$ are small quantities of order $\mathcal{O}(\epsilon)$. Retaining only the leading contributions in the weak-deflection regime, Eq.~(\ref{LE}) simplifies to
\begin{align}\label{Einsteinring}
\theta_\text{E}\simeq \frac{D_{\text{LS}}}{D_\text{OS}}\alpha_\text{D}.
\end{align} Making use of Eq. \eqref{bending4} for the deflection angle and the approximation $\theta\approx b/D_\text{OL}$, we obtain
\begin{eqnarray}\label{bending5}
       \alpha(b)&=&\dfrac{D_{\text{LS}}}{D_\text{OS}}\Bigg[\frac{4(1+16\pi\hat{\rho})M D_\text{OL}}{\theta_E}+\frac{M D_\text{OL}^2}{4\theta_E^2}\Bigg(15\pi M -16a+8\pi\hat{\rho}\Big(\left(-16+9\pi\right)M-8a\nonumber\\&&+8\pi\hat{\rho} M\left(
       60\pi-92\right)\Big)\Bigg)+\frac{M D_\text{OL}^3}{3\theta_E^3}\Bigg(128M^2+6a\left(2a-5\pi M\right)+16\pi\hat{\rho}\Big(\left(208-45\pi\right)M^2\nonumber\\&&+12a\left(4M(\pi-1)+a\right)+16\pi\hat{\rho}\Big(\left(192-54\pi\right)M^2-6aM\left(5\pi-8\right)\nonumber\\&&+16\pi\hat{\rho} M^2\left(196-45\pi\right)\Big)\Big)\Bigg)\Bigg],
   \end{eqnarray}
   \begin{figure}
	\begin{centering}
		    \includegraphics[width=0.5 \textwidth]{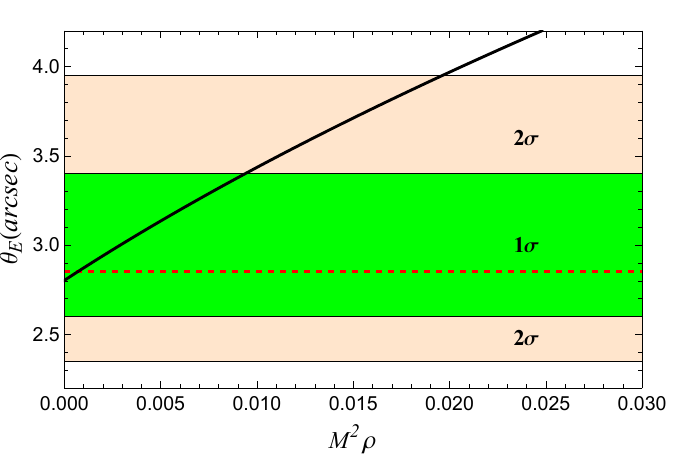}
    		   
	\end{centering}
	\caption{The estimation of angular radius of Einstein ring $\theta_E$ vs DMH parameter $\hat{\rho}$ with the Hubble Space Telescope observations. The uncertainties in $1\sigma$ and $2\sigma$ confidence levels are represented by green and light orange shaded regions, respectively. We kept $a=0.9$}\label{plotWLC}		
\end{figure} By employing numerical methods, we solved Eq. \eqref{bending5} to get the angular size of the Einstein rings, $\theta_E$, by taking galaxy ESO325-G004 as our lens and illustrated the results as a function of DMH parameter $\hat{\rho}$ in Fig.~\ref{plotWLC}. Notably, $\theta_E$ shows a monotonic increment as we increase DMH parameter $\hat{\rho}$. By anlysing the Fig.~\ref{plotWLC}, the DMH parameter $\hat{\rho}$ can be constrained as $0\leq \hat{\rho}\lesssim 0.00939$ within the $1\sigma$ confidence level, and  $0\leq \hat{\rho}\lesssim 0.01963$ within the $2\sigma$ confidence level. Consequently, the weak gravitational lensing signatures of the new axisymmetric black holes immersed in Hernquist DMH offer an additional observational avenue for testing deviations from GR induced by the surrounding dark matter distribution. These effects complement information obtained from strong-lensing phenomena and black hole shadow observations, providing an independent probe of the influence of DMHs on the underlying spacetime geometry.

\section{Conclusion}\label{Sec:Conclusion}

In this work, we investigated the optical properties of a rotating black hole immersed in a Hernquist DMH. The rotating geometry was considered in a Kerr--like form, in which the effects of the surrounding halo were encoded in the radial function $\Delta(r)$ through the density parameter $\rho$. In the appropriate limits, the spacetime reduced to the Kerr geometry when the halo contribution was removed, to the static Hernquist black hole when the rotation parameter was switched off, and to the Schwarzschild solution when both deformations were simultaneously suppressed.

We first analyzed the geodesic structure of the spacetime. By using the conserved quantities associated with stationarity and axial symmetry, we derived the radial equation of motion and the corresponding effective potentials. The behavior of these potentials showed how the rotation parameter $a$ and the Hernquist density parameter $\rho$ modified the allowed regions for particle motion. In the null case, we obtained the condition for photon propagation and studied the radial acceleration at the turning points. The numerical trajectories confirmed that the photon dynamics was strongly affected by the combined influence of frame dragging and the surrounding dark matter distribution.

We then examined the photon region and the shadow cast by the black hole. By exploiting the separability of the Hamilton--Jacobi equation, we derived the critical impact parameters associated with unstable spherical photon orbits and projected them onto the observer's celestial plane. The resulting shadow contours showed two distinct effects. The rotation parameter mainly shifted the shadow and increased its left--right asymmetry, while the Hernquist halo enlarged the photon capture region and increased the apparent size of the shadow. In other words, the parameter $\rho$ acted as an additional source of gravitational attraction for null rays and produced a measurable deviation from the Kerr prediction.

We also confronted the theoretical shadow size with the Event Horizon Telescope measurements of Sgr A$^{\ast}$ and M87$^{\ast}$. By using the area equivalent shadow radius, we obtained upper bounds on the dimensionless halo parameter $\hat{\rho}=M^{2}\rho$. For Sgr A$^{\ast}$, the allowed range was approximately $\hat{\rho}^{\rm max}_{\rm Sgr\,A^{\ast}}\sim (2.7-3.8)\times 10^{-3}$ at $1\sigma$ and $\hat{\rho}^{\rm max}_{\rm Sgr\,A^{\ast}}\sim (4.1-5.2)\times 10^{-3}$ at $2\sigma$. For M87$^{\ast}$, the corresponding bounds were weaker, namely $\hat{\rho}^{\rm max}_{\rm M87^{\ast}}\sim (4.4-5.4)\times 10^{-3}$ at $1\sigma$ and $\hat{\rho}^{\rm max}_{\rm M87^{\ast}}\sim (6.7-7.6)\times 10^{-3}$ at $2\sigma$.

Finally, we studied the gravitational lensing signatures of the rotating Hernquist black hole in both the strong-- and weak--field regimes. In the strong--field regime, the halo contribution shifted the unstable photon orbit and modified the critical impact parameter, thereby changing the logarithmic structure of the deflection angle and the associated relativistic images. In the weak--field regime, the Hernquist contribution appeared already in the leading terms of the bending angle and increased the deviation from the Kerr case as $\rho$ grew. By applying the Einstein--ring data of ESO325-G004, we found the bounds $0\leq \hat{\rho}\lesssim 0.00939$ at $1\sigma$ and $0\leq \hat{\rho}\lesssim 0.01963$ at $2\sigma$.

As a further perspective, it would be worthwhile to investigate the thermodynamic aspects of the present geometry through the optical--mechanical analogy and ensemble-theory methods, following the recent analyses reported in Refs.~\cite{furtado2023thermal,araujo2023thermodynamics,araujo2021bouncing,araujo2023thermodynamical,araujo2022thermal,oliveira2020relativistic}. In addition, the influence of the surrounding Hernquist halo on neutrino oscillations could also be explored, in analogy with recent studies of neutrino propagation in curved backgrounds and modified black hole geometries \cite{araujo2025geodesics,filho2024implications,shi2025influence}.


\section*{Acknowledgments}
\hspace{0.5cm} A.A.A.F. is supported by Conselho Nacional de Desenvolvimento Cient\'{\i}fico e Tecnol\'{o}gico (CNPq) and Fundação de Apoio à Pesquisa do Estado da Paraíba (FAPESQ), project numbers 150223/2025-0 and 1951/2025. N. H is supported by the Conselho Nacional de Desenvolvimento Científico e Tecnológico (CNPq), grant No. 152891/2025-0. N. H. also acknowledges the networking support provided by COST Action CA22113 -- Fundamental challenges in theoretical physics (Theory and Challenges), CA21106 -- COSMIC WISPers in the Dark Universe: Theory, astrophysics and experiments (CosmicWISPers), CA21136 -- Addressing observational tensions in cosmology with systematics and fundamental physics (CosmoVerse), and CA23130 -- Bridging high and low energies in search of quantum gravity (BridgeQG).

\section*{Data Availability Statement}

Data Availability Statement: No Data associated with the manuscript

\bibliographystyle{ieeetr}
\bibliography{main}

\end{document}